%% file: main.tex
\documentclass[twocolumn, modern]{aastex631}

\newcommand{\Var}{\mathrm{Var}}

\usepackage{savesym}
\savesymbol{tablenum}
\usepackage{siunitx}
\restoresymbol{SIX}{tablenum}
\DeclareSIUnit\angstrom{\text {Å}}
\DeclareSIUnit\parsec{pc}
\DeclareSIUnit\deg{deg}
\DeclareSIUnit\mag{mag}
\DeclareSIUnit\erg{erg}

\usepackage{amsmath}
\usepackage{bm}
\usepackage{mathptmx}

\begin{document}

\title{Mid-Infrared Variability in Nearby Galaxies from the MaNGA Sample}

\author[0009-0008-9641-6065]{Aashay Pai}
\affiliation{Center for Cosmology and Particle Physics \\
Department of Physics, New York University, 726 Broadway, New York, NY 10003, USA}

\author[0000-0003-1641-6222]{Michael R.~Blanton}
\affiliation{Center for Cosmology and Particle Physics \\
Department of Physics, New York University, 726 Broadway, New York, NY 10003, USA}

\author[0000-0002-2733-4559]{John Moustakas}
\affiliation{Department of Physics and Astronomy, Siena College, 515 Loudon Road, Loudonville, NY 12211, USA}

\correspondingauthor{Aashay Pai}
\email{aashay.pai@nyu.edu}
\begin{abstract}
We use mid-infrared variability in galaxies to search for active
galactic nuclei (AGN) in the local universe. We use a sample of 10,220
galaxies from the Mapping Nearby Galaxies at APO (MaNGA) survey, part
of the Sloan Digital Sky Survey (SDSS-IV). For each galaxy, we examine
its mid-infrared variability in the  $W2$ $[\SI{4.6}{\um}]$ band
over thirteen years using data from the Wide Infrared Survey Explorer 
(WISE) All-Sky and  Near Earth Objects WISE  (NEOWISE) missions. 
We demonstrate that we can detect variability signatures as small as
about 7\% in the root-mean-square variation of $W2$ flux for the majority of cases. 
Using other AGN signatures of the variable galaxies, such as optical narrow 
lines, optical broad lines, and WISE $W1-W2$ colors, we show that $\sim 75\%$ of 
the variables show these additional AGN signatures, indicating that the 
bulk of these cases are likely to be AGN. We also identify seven 
galaxies that have light-curves characteristic of tidal disruption events.
We present here a publicly available catalog of the light-curve variability 
in $W2$ of these galaxies. 
\end{abstract}

% $W1$ $[\SI{3.4}{\um}]$ and 

\keywords{ Active Galactic Nuclei, MaNGA, WISE}

\section{\label{sec:intro} Introduction}

Active Galactic Nuclei (AGN) are accreting supermassive black holes at
the centers of galaxies, whose radiation across the electromagnetic
spectrum is powered by energy released as matter descends into the
black holes' potential wells.

Emission from the AGN accretion disk, where most of this energy
release occurs, heats the gas and dust surrounding it. Many AGN are
surrounded by a dusty structure, with a size scale of order
$\SI{1}-\SI{10}{\parsec}$, referred to as the ``dust torus''
(\citealt{hickox18a}).  The inner parts of this torus reach up to
about \SI{1500}{\K}, corresponding to the dust sublimation temperature
(at smaller radii from the black hole, the dust is therefore
destroyed). These inner regions are responsible for thermal dust
emission peaking around $\SI{2}-\SI{4}{\micro\m}$ (\citealt{lyu22a}),
with the outer, cooler regions contributing emission out to
\SI{40}{\micro m}.

Closer to the accretion disk than the dusty torus typically lies 
a broad line region (BLR), characterized by high gas velocities 
(1,000s of \SI{}{\km\per\s}) within the gravitational sphere of 
influence of the black hole. Galaxies with observed broad emission 
lines are classified as Seyfert Type 1. However, the BLR can be obscured 
by the dusty torus and in
such cases is unobservable except as polarized scattered light
(\citealt{antonucci83a}). These AGN are
typically still observable from emission lines emanating from the
Narrow Line Region, which extends to 100s of pc and sometimes several
kpc (\citealt{hickox18a}). These AGN are classified as Seyfert Type 2.

Their mid-IR emission make AGN spectra stand out from purely
star-forming galaxies. The former have narrow emission lines
interspersed between a smooth continuum, whereas the latter show
prominent Polycyclic Aromatic Hydrocarbon (PAH; \citealt{smith07a}) lines and
substantially cooler thermal dust emission instead. 
%PAHs are large molecules (a few $\AA$ in diameter) containing $\sim 100$ carbon atoms. 
This difference makes the
Wide Infrared Survey Explorer (WISE; introduced in
Section~\ref{sec:wise}) a natural candidate to be used for identifying
AGN.  \cite{stern12a} introduced criteria in WISE $W1$ and $W2$ fluxes
to detect a mix of Type 1 and Type 2 AGN.

\begin{figure*}[t!]
    \centering
    \includegraphics[width=0.94\textwidth]{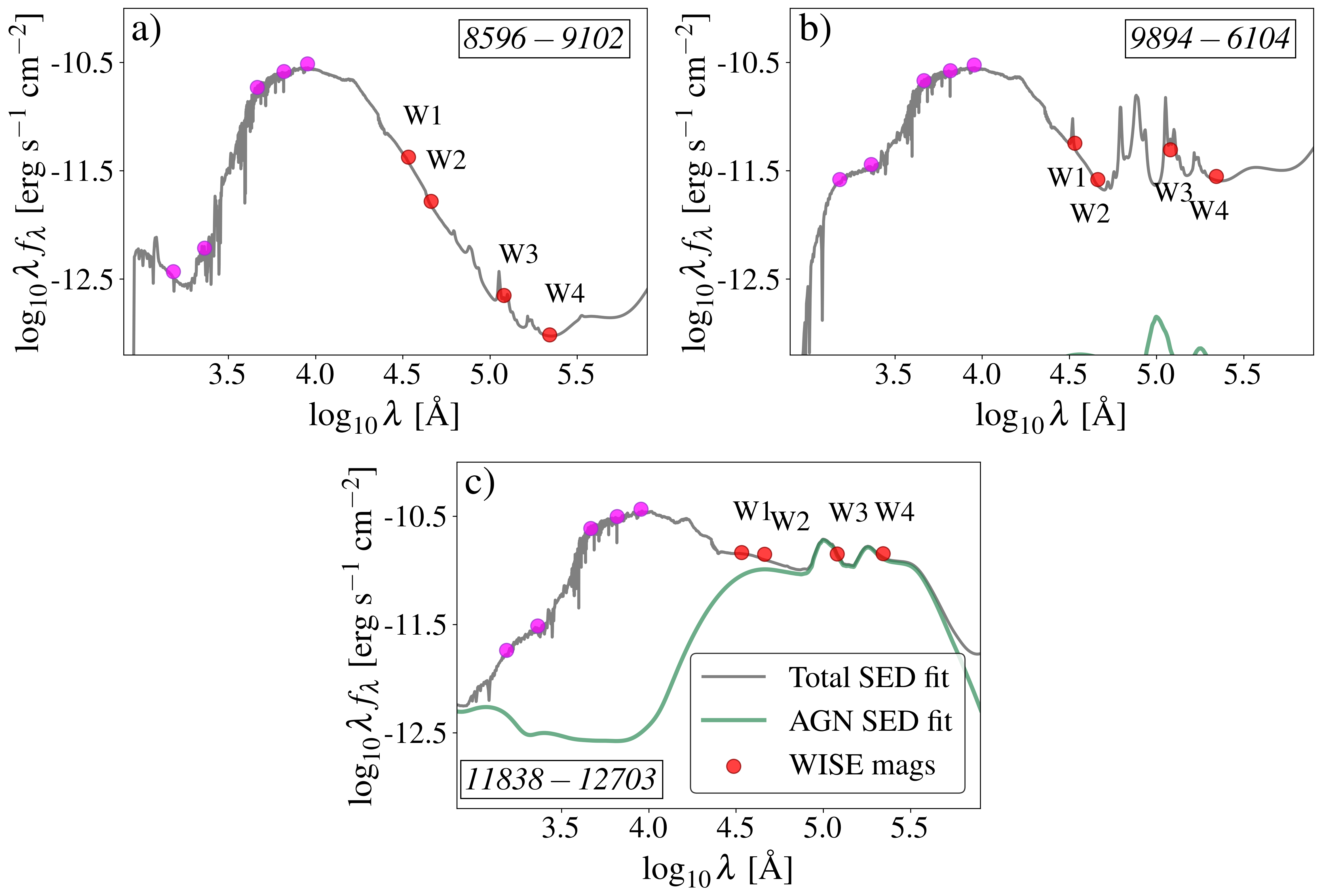}
    \caption{MaNGA SED fits on (a) quiescent, (b) star-forming, and
      (c) AGN galaxies. The grey lines are the total SED fit to broad band photometry, the green lines are the estimated AGN component. The fuchsia points are the GALEX NUV \& FUV and the LS $grz$ bands (from left to right). The red points are the four WISE magnitudes - $W1, W2, W3$ and $W4$. Each panel is labelled by its MaNGA identifier plate-IFU. Only the galaxy dominated by the AGN component in mid-IR demonstrates an increase
      in $W2$ flux, therefore contributing to a redder $W1-W2$ color.}
    \label{fig:spsmodels}
\end{figure*}

As an illustration, Figure~\ref{fig:spsmodels} shows Spectral Energy Distribution (SED) fits to galaxy 
broad band photometry for three MaNGA galaxies. These fits are to ultraviolet
far- and near-ultraviolet bands from the Galaxy Evolution Explorer (GALEX; \citealt{morrissey07a}),
the DESI Legacy Imaging Surveys (LS; \citealt{dey19a}), and WISE photometry.
The model contains two different contributions,
fit simultaneously. 
The first is a stellar population synthesis fit, using a nonnegative linear 
combination of a broad range of templates
from the Flexible Stellar Population Synthesis code (FSPS; \citealt{conroy10a}); these
templates include dust attenuation and reemission. The second uses templates from 
the AGN, using the CLUMPY models of \citet{nenkova08a}. The black line is the 
total SED, and the green is the AGN contribution.

Each spectrum represents a different type of galaxy---quiescent, star-forming or AGN. The WISE bands are labeled in the mid-infrared wavelengths. 
Panel (a) shows a quiescent galaxy with a steep, very blue mid-infrared
spectrum dominated by the Rayleigh-Jeans tails of the stellar emission. 
Galaxies like this one have correspondingly low $W3$ and $W4$ fluxes, 
because they do not 
contain large amounts of warm dust, the main contributor to this emission. 
Panel (b) shows a star-forming galaxy, with high $W3$ and $W4$ due to the 
presence of the warm dust and PAHs associated with star formation
The ratio of $W1$ to $W2$ fluxes for both quiescent and star-forming galaxies 
are largely unperturbed by the warm dust and PAHs, and these galaxies 
typically have bluer $W1-W2$ colors.  Panel (c) shows a galaxy with an
AGN in which the SMBH heats the dust to sublimation temperatures, causing 
an increase in the $W2$ emission from this hot dust, thereby changing the 
$W1$ to $W2$ ratio, and causing a redder $W1-W2$ color.

In this paper we seek to identify variable AGN. Whereas longer IR 
wavelengths are less prone to variation, near IR wavelengths around 
$1$--$\SI{5}{\micro\m}$ vary on human timescales, because this emission is caused 
by the dust reverberation response due to variations in the accretion disk, 
as confirmed by the time lag between observed IR and UV fluxes (\citealt{lyu22a}). 
Based on the above discussion of Figure \ref{fig:spsmodels}, the variability in 
the amplitude of the AGN will  be fractionally larger in
$W2$ than $W1$. For many star forming galaxies the fractional $W2$
variation will also exceed $W3$ or $W4$. Since WISE observed in $W1$ and $W2$ longer than $W3$ and $W4$, we will concentrate on the variability of $W2$ here.

% \cite{assef13a} built upon the previous study and extended the region
% of interest to the bigger $\SI{9} {\square\deg}$ field of NOAO Deep
% Wide-Field Survey Bo\"otes field. They found that the color cut
% proposed above was not as reliable when considering faint sources, and
% in turn, defined a modified criterion for selecting AGN at fainter
% fluxes using the $W1-W2$ color and $W2$ magnitude from WISE.
 
With these motivations in mind, we create a WISE catalog using the
MaNGA sample (introduced in Section \ref{sec:manga}), by analyzing the
variability of the WISE All-Sky and NEOWISE observations of these
galaxies. We further define a new variability criterion to identify
AGN in the MaNGA sample. In Section \ref{sec:galaxysel} we give 
descriptions of MaNGA and WISE. In Section \ref{sec:overview} we give a broad overview of our methodology to determine variability, in Section \ref{sec:catcontents} we describe 
the calculations performed to determine variability and the contents of the catalog, 
in Section \ref{sec:characterizing_var} we define what objects we consider 
to be variable, and in Section \ref{sec:properties} we verify our variable 
sample by looking at their optical narrow line, optical broad line and 
$W1-W2$ colors. We show MaNGA images and spectra of a few variable galaxies in Section \ref{sec:var_examples}. Section \ref{sec:summary} summarizes our work and Section \ref{sec:dataavail} explains how the catalog can be accessed.

All magnitudes are on the AB system (\citealt{oke70a}) throughout. We have converted the
Vega-relative magnitudes published by WISE using the {\tt kcorrect 5.1.2} Python package (\citealt{blanton07b}).

\section{\label{sec:methods} Methods} 

\subsection{Galaxy Selection} \label{sec:galaxysel}

The galaxies in this catalog are selected from the Mapping Nearby
Galaxies at APO (MaNGA; \citealt{bundy15a}) component of the Sloan
Digital Sky Survey IV (SDSS-IV; \citealt{blanton17a}). We match these
galaxies to the Wide Infrared Survey Explorer's (WISE;
\citealt{wright10a}) WISE All-Sky and Near Earth Objects WISE
(NEOWISE; \citealt{mainzer14a}) missions.
%We further filtered the queried galaxies to eliminate anomalous sources. 

\subsubsection{MaNGA Overview} \label{sec:manga}

MaNGA is an integral field survey that observed 11,273 objects using the Sloan 
Foundation Telescope at Apache Point Observatory, New Mexico as part of SDSS-IV. 
It used the Baryon Oscillation Spectroscopic Survey (BOSS) spectrographs
connected to 17 optical fiber bundles (\citealt{drory15a, smee13a}).

The sample contains all types of nearby galaxies ($z \leq 0.15$) with stellar 
masses between $10^9$ to $10^{11}$ $M_\odot$,  divided into Primary, Secondary, 
and Color-Enhanced samples. Using these observations, the 
Data Reduction Pipeline built data cubes by extracting spectra from the 
fibers (\citealt{law15a, law16a}). The Data Analysis Pipeline then further 
processed the data to extract velocity dispersion and emission line 
flux measurements (\citealt{westfall19a}). 

\citet{sanchez22a} analyzed the MaNGA datacubes using the {\tt
  pyPipe3D} data pipeline, to produce the Pipe3D
Value Added Catalog, which we use here. This catalog contains various
measurements regarding the stellar populations, star 
formation histories, emission lines, and other properties of galaxies.

We create and analyze images for GALEX far- and near-ultraviolet, 
LS $grz$, and WISE bands. We calculate elliptical aperture
photometry for all MaNGA galaxies, which we use here for mid-infrared
$W1-W2$ color measurements.
  
\subsubsection{\label{sec:wise} WISE Overview} 

WISE is a NASA Explorer launched in 2009 and put into hibernation in
2011. WISE collected data using a 40-cm cryogenically-cooled
telescope which contained four mid-infrared detectors and imaged the
entire sky in band passes at $3.4, 4.6, 12$, and $\SI{22}{\um}$,
referred to as $W1, W2, W3$, and $W4$. These band passes have a
resolution of $6.1'', 6.4'', 6.5''$, and $12.0''$
(\citealt{wright10a}).

The WISE All-Sky survey contains data collected during the full
cryogenic mission phase, which took place from January 2010 through
August 2010. The processed catalog contained photometric observations
of over 563 million detected objects.

WISE was brought out of hibernation in 2013 for the NEOWISE
mission (\citealt{mainzer14a}). The WISE Moving Object Pipeline System searches for
near-Earth objects, comets, and small planets. The NEOWISE mission
also processes observations of other WISE objects through the standard
WISE pipeline used for the All-Sky survey.

\subsection{\label{sec:overview} Overview of Methodology}

We calculate $W2$ variability because, as explained above, it is a good 
indicator of AGN activity. To do so, we search for objects with a variance 
in their observed fluxes that exceed the variance expected from observational
uncertainties. We use the individual WISE All-sky and NEOWISE $W2$ observations 
binned into ``epochs,'' which WISE observed on a six-month  cadence. 
We then use these $W2$ light-curves to calculate the observed variance across 
all epochs of the mean magnitudes per epoch  (Equation \ref{eq:obs_var_all_mags}). 
We compare this quantity to the variance in the mean magnitude per epoch 
expected from an empirical estimate of the observational uncertainties 
(Equation \ref{eq:exp_var_all_mags}). This procedure estimates 
the $W2$ variability relative to expectations. To determine which objects are 
reliably classifiable as variable, we perform a Monte Carlo simulation to identify
how much larger than expectations the variance must be to ensure a pure sample
of variables. We define objects with higher variability than expected in this sense
as ``variable in the mid-IR." 

\subsection{\label{sec:catcontents} Quantifying $W2$ Variability} 

We combine the WISE All-Sky and NEOWISE observations for each object
to quantify the variability of these objects between January 2010 and
December 2022. We use data from the single source ``L1b'' tables,
specifically the {\tt w1mpro} and {\tt w2mpro} columns, which contain
magnitudes measured with profile-fitting photometry. During each epoch
of observation, WISE made an average of 15 observations of each object
over the span of a day. There are two epochs during each year for
which WISE was active.  
We query the WISE database with a tolerance of 2 arcseconds in right ascension and declination, and then further filter out all matches that have uncertainties in RA and Dec of more than 1 arcsecond. We also require that the overall frameset quality score flag ({\tt qual\_frame})---which can be 0, 5 or 10, with 10 being the best quality---be nonzero. Finally, we filter out all observations that have been flagged with a contamination and confusion flag ({\tt cc\_flags}), which identifies diffraction spikes, persistence, anneal-resistant latent images, halos, optical ghosts, long-term latent images and glints. We only use galaxies that have detections in 4 or more epochs (using {\tt epoch\_flag}, see Table~\ref{tab:cat_defs}).

We calculate the following statistical
quantities to determine the variability of the galaxies in the sample. All quantities are calculated using the observations stored in the {\tt w1mpro} and {\tt w2mpro} columns and the estimated errors in the {\tt w2sigmpro} and {\tt w2sigmpro} columns. In the notation below, $j$ indexes each epoch and $i$ indexes each observation within an epoch.

\begin{itemize}
    \item Number of good observations at each epoch $n_j$.
    \item Number of epochs per object $N$.
    \item Mean $W2$ magnitude of each epoch.
    \begin{equation}
    \overline{W2}_j \equiv \frac{1}{n_j}\sum_{i=1}^{n_j}(W2)_{ij}
    \end{equation}  
    
    \item Expected variance within each epoch (based on catalog errors): expected variance in the magnitudes within each epoch estimated from the reported catalog uncertainties.
    \begin{equation}
        \sigma_{{\rm ep}, j}^2 \equiv \frac{1}{n_j}\sum_{i = 1}^{n_j}\sigma_{ij}^2
    \end{equation}
    
    \item Observed variance within each epoch (based on within-epoch magnitudes): variance in the observed W2 magnitudes per epoch.
    \begin{equation}
    \label{eq:exp_var_per_epoch_mags}
    \Var{(W2_{{\rm ep}, j})} \equiv \frac{1}{n_j-1}\sum_{i = 1}^{n_j}
      \left(W2_{ij} - \overline{W2}_j\right)^2 
    \end{equation}

\item Expected variance in the mean within each epoch (based on catalog errors).
\begin{equation}
    \sigma_{j}^2 \equiv \frac{\sigma_{{\rm ep}, j}^2}{n_j}     
\end{equation}

\item Expected variance in the mean within each epoch (based on within-epoch magnitudes): 
\begin{equation}
    \label{eq:exp_var_mean_per_epoch_mags}
    \varsigma_j^2 \equiv \frac{\Var{(W2_{{\rm ep}, j})}}{n_j}
\end{equation}

\item Error in mean per epoch (using the quantity defined in Equation~\ref{eq:exp_var_mean_per_epoch_mags}).
    \begin{equation}
        \sigma_{{\rm \overline{W2}}, j} \equiv \varsigma_j
    \end{equation}
\item Observed variance across all epochs: the variance of the mean $W2$ magnitudes of each epoch (Equation \ref{eq:obs_var_all_mags}).
First, we define the mean of the magnitude over all epochs:
\begin{equation}
    \label{eq:mean_mag_all_epochs}
    \mu(\overline{W2}) \equiv \sum_{j = 1}^N\frac{\overline{W2}_j}{N}
\end{equation}
and we use it to calculate the observed variance:
\begin{equation}
    \label{eq:obs_var_all_mags}
     \Var{(W2)} \equiv \frac{\sum_{j = 1}^{N}\left(\overline{W2}_j - \mu(\overline{W2})\right)^2}{N-1}
\end{equation}
\item Expected variance across all epochs (based on catalog sigma): the epoch-to-epoch variance in the mean magnitude calculated using the estimated errors \\
\begin{equation}
    \label{eq:exp_var_all_sigma}
   \sigma^2 \equiv \sum_{j=1}^N\frac{\sigma_j^2}{N}
\end{equation}

\item Expected variance across all epochs (based on within-epoch
  variances): the epoch-to-epoch variance in the mean magnitude
  calculated using the per epoch variance in the observed magnitude,
\begin{equation}
    \label{eq:exp_var_all_mags}
    \varsigma^2 \equiv \sum_{j=1}^N\frac{\varsigma_j^2}{N}
\end{equation}
\end{itemize}

\begin{figure}[h!]
    \centering
    \includegraphics[width=\columnwidth]{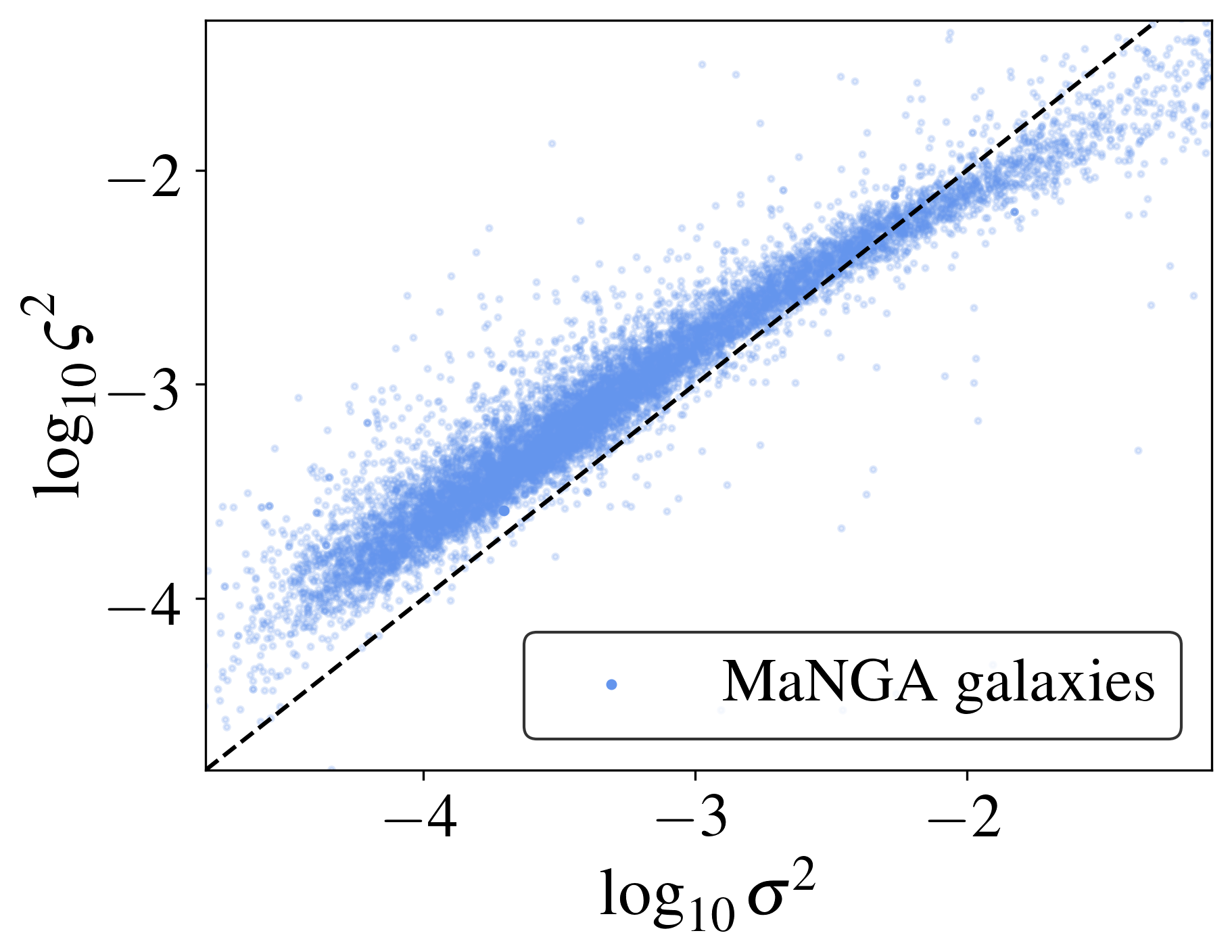}
    \caption{Comparing the expected variance derived from within-epoch variances (Equation~\ref{eq:exp_var_all_mags})
     to the expected variance derived from within-epoch uncertainties (Equation~\ref{eq:exp_var_all_sigma}). 81\% of the MaNGA sample lies above the one-to-one line ($\varsigma^2 > \sigma^2$). We therefore conservatively use $\varsigma^2$ as our true errors.}
     \label{fig:errs_vs_mags}
\end{figure}

Figure~\ref{fig:errs_vs_mags} compares the expected variance across all epochs derived from within-epoch variances ($\varsigma^2$ in Equation~\ref{eq:exp_var_all_mags}) to the expected variance across all epochs derived from reported catalog uncertainties ($\sigma^2$ in Equation~\ref{eq:exp_var_all_sigma}). 
There is substantial disagreement between these estimates, and for 
around 81\% of our galaxies, $\sigma^2$ is much larger. 

We do not expect that these galaxies show substantial variability on hour or 
day long time scales. In any case, to explain the difference, the variability
would have to correlate well with the expected uncertainties, which seems
implausible. Therefore this result appears to show that the reported 
uncertainties are unreliable and do not reflect the true uncertainties. 
Therefore, we use Equation~\ref{eq:exp_var_all_mags} instead of  
Equation~\ref{eq:exp_var_all_sigma} for our analysis in the subsequent 
sections.

\subsection{\label{sec:characterizing_var}Identifying Variable Sources}

\begin{figure*}[t!]
    \centering
    \includegraphics[width=\textwidth]{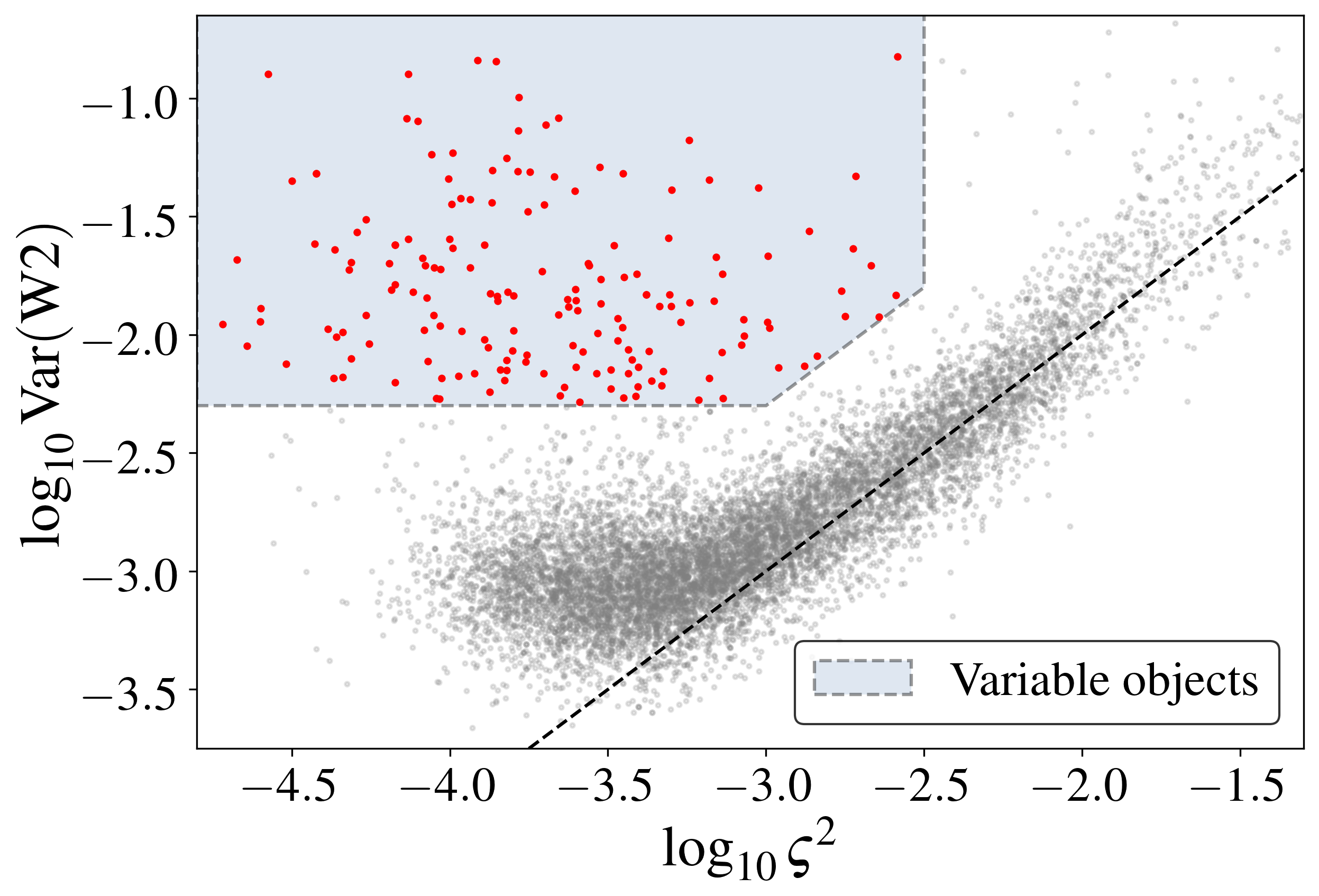}
    \caption{Observed $W2$ variance across all epochs
      (Equation \ref{eq:obs_var_all_mags}) versus the expected $W2$ variance
      across all epochs calculated from within-epoch variances
      (Equation \ref{eq:exp_var_all_mags}) for MaNGA galaxies. 
      Galaxies in red
      are defined to be ``variable in the mid-IR'' in
      Sec.~\ref{sec:characterizing_var}.}
    \label{fig:var_objs}
\end{figure*}

We determined the variability of the MaNGA galaxies using the
epoch-to-epoch variability calculations performed in Sec.~
\ref{sec:catcontents}. Figure \ref{fig:var_objs} compares the observed
variance across all epochs (Equation \ref{eq:obs_var_all_mags}) versus the
expected variance across all epochs based on within-epoch variances
(Equation \ref{eq:exp_var_all_mags}).
Most of the galaxies show a one-to-one
correspondence between the observed and expected variances, and we
classify these galaxies as non-variable. They show slightly higher
observed variance then expected, which is presumably due to systematic
errors arising between epochs that are not captured in the
within-epoch variance. This additional variance appears to become a little
stronger at higher expected variance values (in the
upper right) and at the lowest expected variances (the ``floor'' in
the the observed variance below $\log_{10} \mathcal{\varsigma}^2 \sim
-3.0$). The floor
presumably arises due to calibration-related systematics across
epochs.

Galaxies in the top-left of this diagram exhibit variability over epochs that 
is much higher than that which is expected of them due to noise, and we classify these objects 
as variable in the mid-IR. We have a choice in how we design our classification
scheme; we can set a high threshold for variability, which will be detectable for 
a larger fraction of the objects, or a lower threshold, which is only detectable 
for the highest signal-to-noise objects. Figure \ref{fig:var_objs} shows our 
default choice as the shaded region, which is defined in Equation \ref{eq:shadedregion}.
\begin{equation}
\label{eq:shadedregion}
    \begin{split}
    & \log_{10}\Var{(W2)} > -2.3\\ 
   &  \log_{10}\varsigma^2 < -2.5\\ 
   & \log_{10}\Var{(W2)} > \log_{10}\varsigma^2 + 0.7
    \end{split}
\end{equation}

We determined this region by performing a Monte Carlo simulation consisting of 100,000 realizations to
estimate the expected distribution of $\Var{(W2)}$ in the
hypothetical case that none of the objects are variable.  We
bin the data by $\log_{10} \varsigma^2$, and compute the mean
$\log_{10}\Var{(W2)}$ per bin (which we assume to be the
true variance in the measurements of a non-varying object due to
observational effects). We then randomly sample a galaxy, find the
corresponding mean $\log_{10}\Var{(W2)}$ (based on its bin),
and assume individual epochs will be distributed as a Gaussian with a
variance $\Var{(W2)}$. We sample 25 points from this
Gaussian (which is the average number of epochs for all the galaxies)
and recompute the $\log_{10}\Var{(W2)}$ with this Monte
Carlo data. 

This procedure results in a distribution of $\log_{10}\Var{(W2)}$ 
as a function of $\log_{10} \varsigma^2$. This distribution is actually
somewhat tighter around the data's mean in each bin than the data is.
We therefore apply a fudge factor of $1.25$ to inflate the width
of the Monte Carlo distribution around the mean. We set this factor 
to match the negative tail of the distribution; the origin
of this factor is not entirely clear, but it presumably arises from
non-Gaussianity in the real errors and the fact that some fraction of
galaxies were observed fewer than 25 times.

\begin{figure}[h!]
    \centering
    \includegraphics[width=\columnwidth]{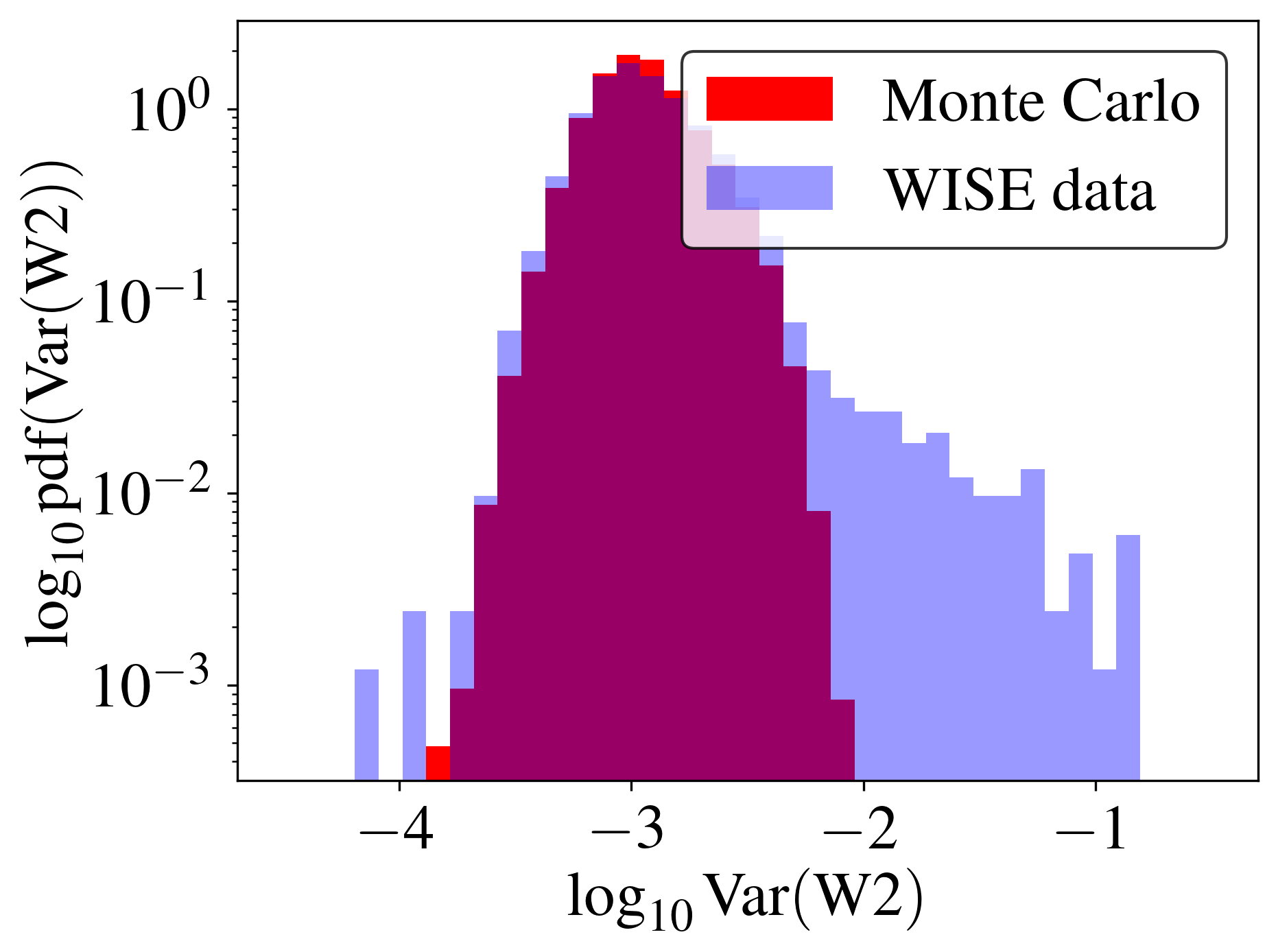}
    \caption{Distribution of predicted points using Monte Carlo
      simulations vs. the observed $\log_{10} \Var{(W2)}$,
      only using points with $\log_{10} \varsigma^2 < -2.5$.}
    \label{fig:montecarlo}
\end{figure}

Fig.~\ref{fig:montecarlo} shows the distribution of $\log_{10}
\Var{(W2)}$ predicted by the Monte Carlo simulations
(including the fudge factor) and the values calculated using WISE
data, for the data with $\log_{10} \varsigma^2 < -2.5$. The 
tail at high $\log_{10} \Var{(W2)}$ contains the identified variable
objects.

Our procedure produces a distribution  that closely follows the 
non-variable object distribution seen in Fig.~\ref{fig:var_objs}.
Using a Monte
Carlo realization with roughly ten times the number of samples as 
we have galaxies, we define the shaded region in Fig.~\ref{fig:var_objs}
conservatively, such that the region contains zero samples from the
Monte Carlo simulation.

\begin{figure*}[]
    \begin{center}
        \includegraphics[width=0.9\textwidth]{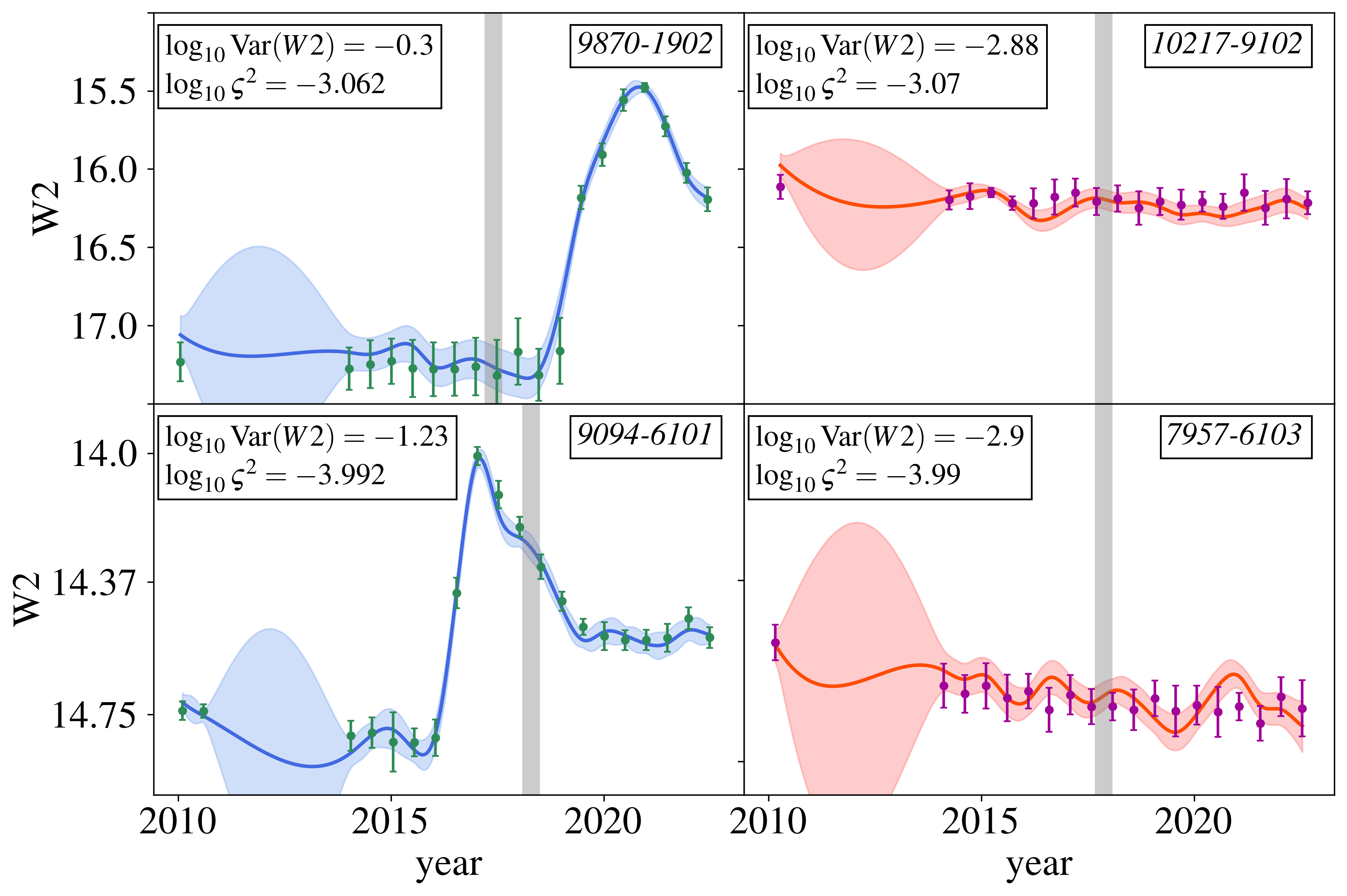}
         \caption{Light-curves of galaxies that are variable in the mid-IR (in blue), and galaxies that are not classified as variable in the mid-IR (in orange). Each panel is labelled by its MaNGA plate-IFU (top right) and its location in our variability space given by $\log_{10}\Var{(W2)}$ and $\log_{10}\varsigma^2$ (top left). The grey band indicates when the object was observed by MaNGA.}
        \label{fig:lightcurves_var_nonvar}
    \end{center}
\end{figure*}

This definition leads to 170 galaxies that are identified as 
variable in the  mid-IR. These objects are tagged with a variability flag ({\tt var\_flag}). Figure~\ref{fig:lightcurves_var_nonvar} shows examples of $W2$ light-curves of two galaxies that we classify as variable in the mid-IR (in blue) and two that have comparable $\varsigma^2$, but are not identified as variable in the mid-IR (in orange). The individual points are WISE observations binned into epochs, and the lines are Gaussian process fits to those points. The Gaussian process regression was done using the {\tt scikit-learn} (\citealt{scikit-learn}) Python package using a Matern-3/2 kernel, with a hyperparameter length scale of $\sim 15$ years. The GP fit has only been used to visualize the data.

We identify 15 galaxies that fall in 
our variability cut but either have data for only a few epochs or 
whose changes in magnitude exist only over one epoch, making 
the variability questionable. We tag these 15 objects with a visual 
inspection flag ({\tt vi\_flag}, as described in 
Table~\ref{tab:cat_defs}).

% \begin{figure}[h]
%     \begin{center}
%         \includegraphics[width=\columnwidth]{var_magcut_vertical.png}
%         \caption{\textbf{(a)} compares $\log_{10} \varsigma^2$ vs. $W2$, with the black line plotting the vertical edge of the variable cut in \textbf{(b)}. The galaxies in blue have $W2<16.5$ and $\log_{10}\varsigma^2<-2.5$. \textbf{(b)} show the population with $W2<16.5$. It is also labelled with 'count', which is the number of variable objects that remain after the magnitude cut.}
%         \label{fig:manga_var_magcut}
%     \end{center}
% \end{figure}

\begin{figure*}[]
    \begin{center}
        \includegraphics[width=\textwidth]{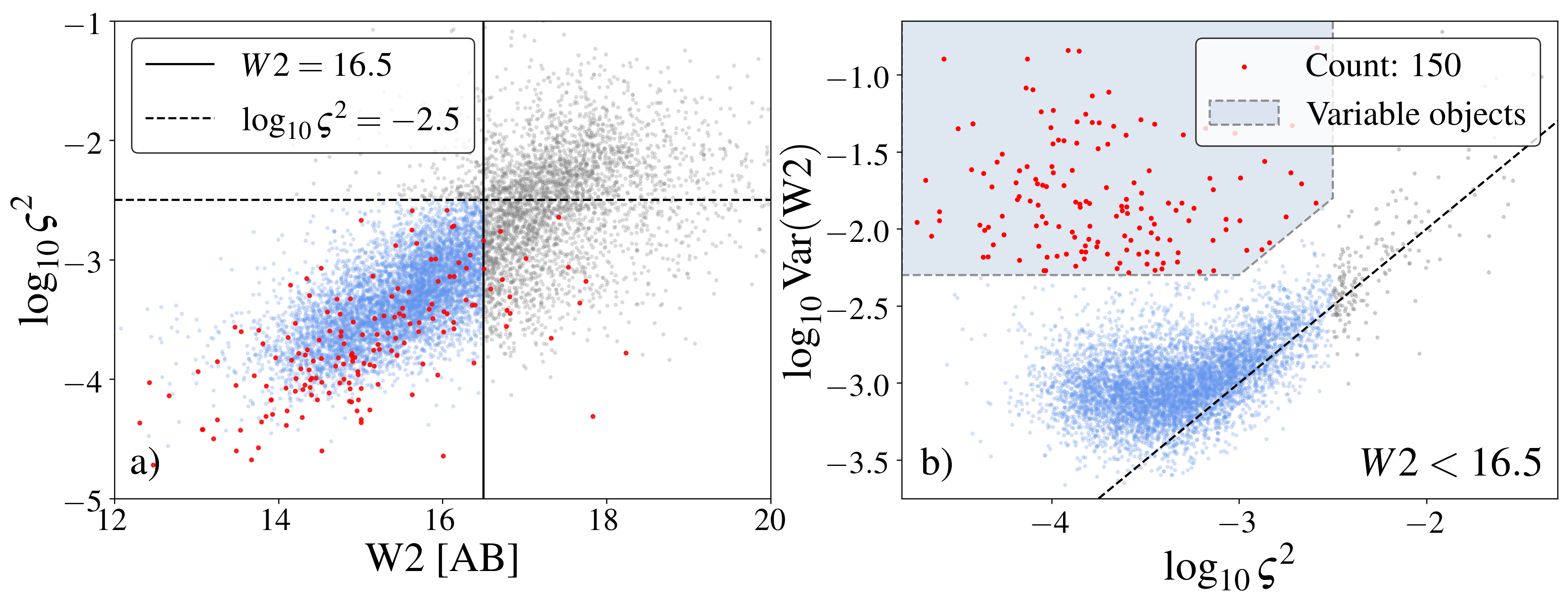}
         \caption{\textbf{(a)} compares $\log_{10} \varsigma^2$ vs. $W2$, with the black line plotting the vertical edge of the variable cut in \textbf{(b)}. The galaxies in blue have $W2<16.5$ and $\log_{10}\varsigma^2<-2.5$. \textbf{(b)} show the population with $W2<16.5$. It is also labelled with 'count', which is the number of variable objects that remain after the magnitude cut.}
        \label{fig:manga_var_magcut}
    \end{center}
\end{figure*}

The expected variance $\varsigma^2$ is primarily determined by the mean $W2$ flux. We therefore look for an approximate $W2$ magnitude above which our variability determination is complete.
Panel (a) of Figure \ref{fig:manga_var_magcut} shows
$\log_{10}\varsigma^2$ and $W2$  for each galaxy. The vertical line 
at $\log_{10}\varsigma^2 = -2.5$  corresponds to the vertical boundary 
of the variable region in  Figure~\ref{fig:var_objs}. The horizontal dashed line is at $W2 = 16.5$; around 65\% of the galaxies have $W2$ brighter than 16.5. In this subsample (highlighted in blue), 97\% of the galaxies have $\log_{10}\varsigma^2<-2.5$.
Panel (b) of Figure~\ref{fig:manga_var_magcut} is similar to Figure 
\ref{fig:var_objs} but only shows the galaxies with $W2<16.5$. We can 
reliably identify variability greater than $\log_{10}\Var{(W2)} > -2.3$ if the galaxy has $W2<16.5$, and our catalog is complete subject to these constraints. 

\subsection{\label{sec:properties} Properties of the Variable Galaxies}

In this section, we examine the source of the variability of the galaxies 
in our catalog by using other well-known methods of identifying AGN.
We consider optical narrow and broad lines in MaNGA spectra and the 
$W1-W2$ color of these galaxies. We also check if our transients match identified Tidal Disruption Events (TDEs) by looking a three catalogs (\citealt{masterson24a}, \citealt{wang12a} \& \citealt{callow24a}). Finally, we check if our variable objects have matches to Mid-IR Outbursts in Nearby Galaxies (MIRONGs; \citealt{jiang21a}) catalog. 

\input{classification}
Table~\ref{tab:classification} summarizes the fraction of variable objects that fall into other classification techniques and types of AGN. 
$N_{\rm var} = 170$ is the total number of variable objects from our catalog. $N_{\rm cat}$ is the number of objects in the catalog being matched to, and $N$ is the number of variable objects from our catalog that are also found in the other catalog. 

\subsubsection{ \label{sec:bl_agn} Broad Line AGN}

\begin{figure}[h!]
    \begin{center}
        \includegraphics[width=\columnwidth]{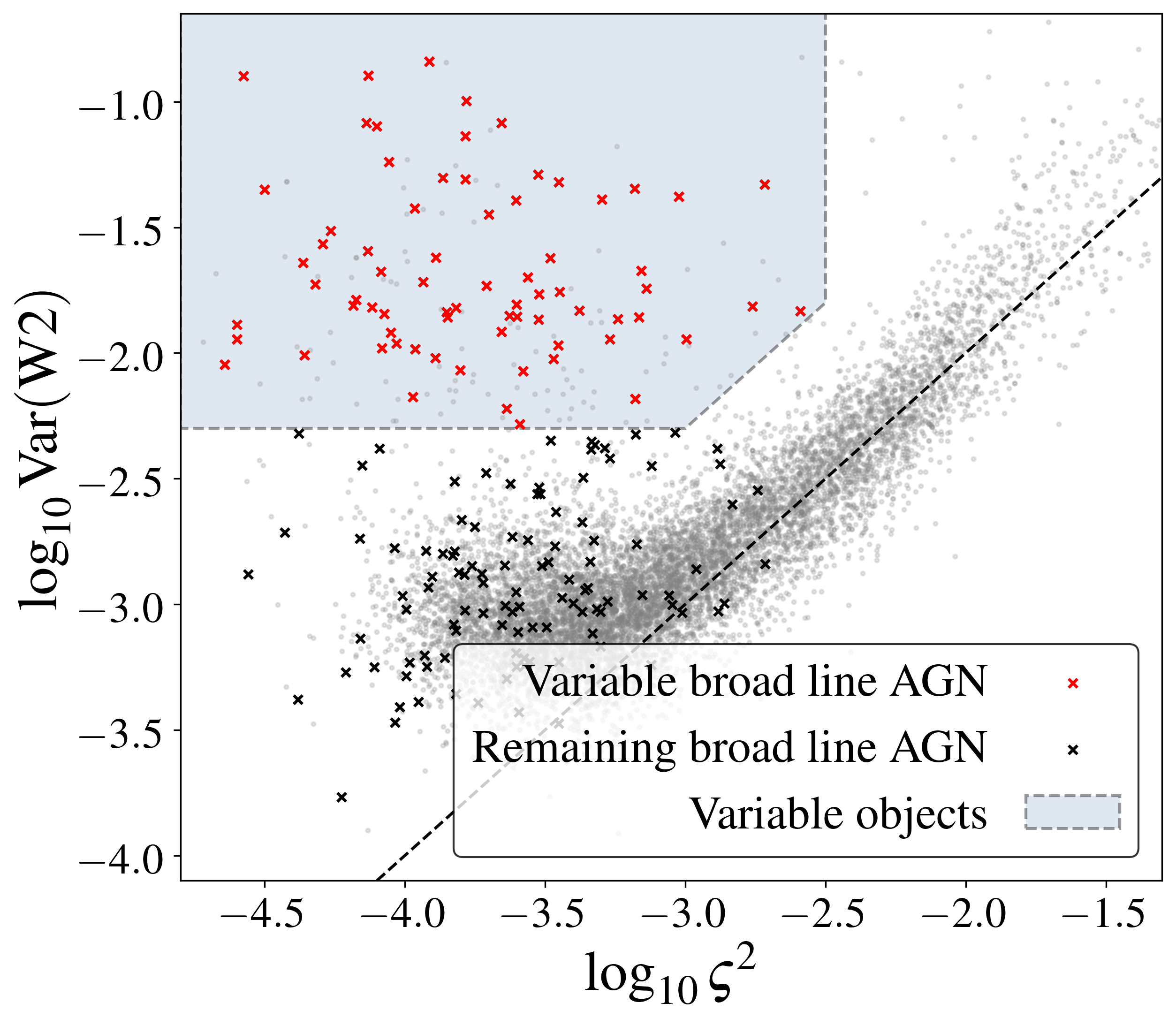}
        \caption{Similar to Figure \ref{fig:var_objs}, showing where broad line AGN lie in the 
        space of W2 variability. The grey points are 
        all of the MaNGA galaxies. The red and black crosses are broad line AGN identified in 
        the broad line catalog of \citet{fu23a}. The red crosses are the broad line AGN that we 
        identify as variable in the mid-infrared, and the black are the broad line AGN that
        we do not identify as variable.}
        \label{fig:var_bl_objs}
    \end{center}
\end{figure}

To examine the overlap between our detected variable galaxies and the detected 
Type 1 AGN population, we use the catalog from \citet{fu23a} of broad-line AGN 
from the MaNGA sample (hereafter called the broad line catalog). 
Figure~\ref{fig:var_bl_objs} shows the distribution of broad line AGN in our variability 
plot. 

Around 42\% of our variable objects have matches in the broad line catalog,
and around 39\% of the broad line AGN are identified as variable by us.
All of the galaxies with broad line AGN have $W2$ fluxes bright enough that we 
expect to detect variability with a log variance (in mag$^2$ units)
greater than about $-2.3$. This variability limit is relative to the total
$W2$ flux (stellar plus AGN).

\subsubsection{Optical Narrow Line AGN}

Optical narrow-line diagnostic diagrams are an effective tool for understanding 
and separating different ionizing sources in galaxies. Baldwin, Philips, \& Terlevich (BPT) 
diagrams (\citealt{baldwin81a}) and Veilleux \& Osterbrock (VO) diagrams 
(\citealt{veilleux87a}) use a combination of $\text{[O III]/H}\beta$, $\text{[N II]/H}
\alpha$ and $\text{[S II]/H}\alpha$ line ratios (hereafter referred to as R3, N2 and S2) to 
differentiate between star-forming, AGN host, and low ionization emission region galaxies 
(LINERs). Specifically, BPT diagrams compare R3 to N2 and VO diagrams compare R3 to S2. 
\citet{jiyan20a} presents a reprojection of the N2, S2 and R3 line ratio space, 
claiming that this reprojection, defined in Equation \ref{eq:jiyan_space}, provides a cleaner
classification than either BPT or VO individually:
\begin{equation}
\label{eq:jiyan_space}
    \begin{split}
    & P1 =~~~0.63 N2 + 0.51 S2 + 0.59 R3 \\ 
   & P2 = -0.63 N2 + 0.78 S2 \\ 
   & P3 = -0.46 N2 - 0.37 S2 + 0.81 R3
    \end{split}
\end{equation}

\begin{figure}[h]
    \centering
    \includegraphics[width=\columnwidth]{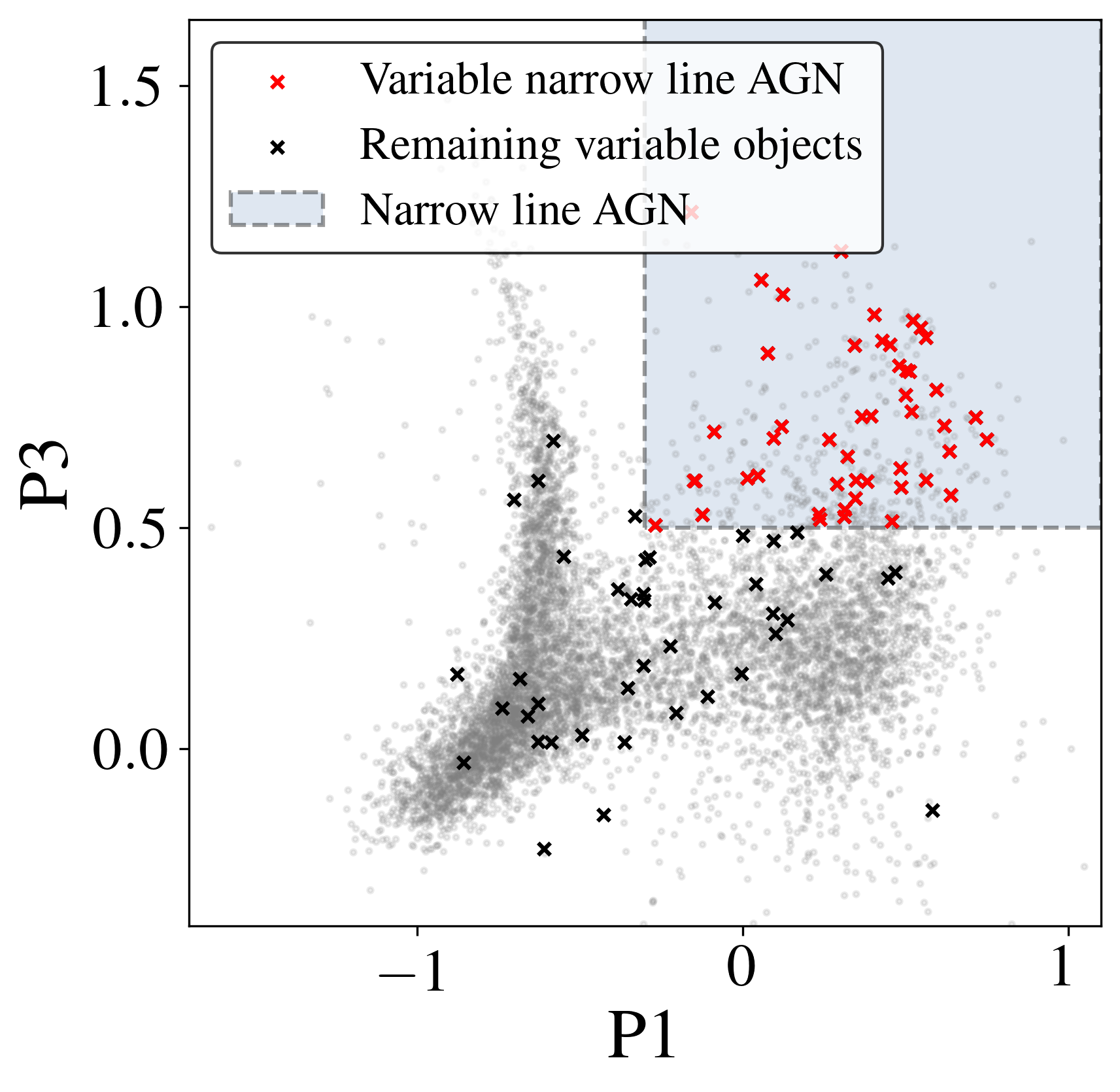}
    \caption{Depicting where our variable galaxies lie in the P1 vs. P3 space, with the grey points being the MaNGA galaxies. The red crosses are variable objects from our catalog that are identified as narrow line AGN based on the criterion from \citet{jiyan20a}. The black crosses are the remaining objects that we identify to be variable in the mid-infrared.}
    \label{fig:p13}
\end{figure}

Figure~\ref{fig:p13} shows the distribution of MaNGA galaxies plotted in the narrow line 
space defined above. 
We exclude known broad line AGN from this plot using the 
broad line catalog from \ref{sec:bl_agn}, because the DAP 
measurements of narrow 
line emission are imperfect in many such cases. 
We classify galaxies with $P1>-0.3$ and $P3>0.5$ as narrow line 
AGN, and it can be seen that roughly half of the non-broad-line variable objects lie within these bounds. 
Around 28\% of all of the variable objects, and around 49\% of the non-broad-line variable objects,
are identified as narrow line AGN. 

\subsubsection{\label{sec:w12} W1-W2 color}

\begin{figure}[h]
    \centering
    \includegraphics[width=\columnwidth]{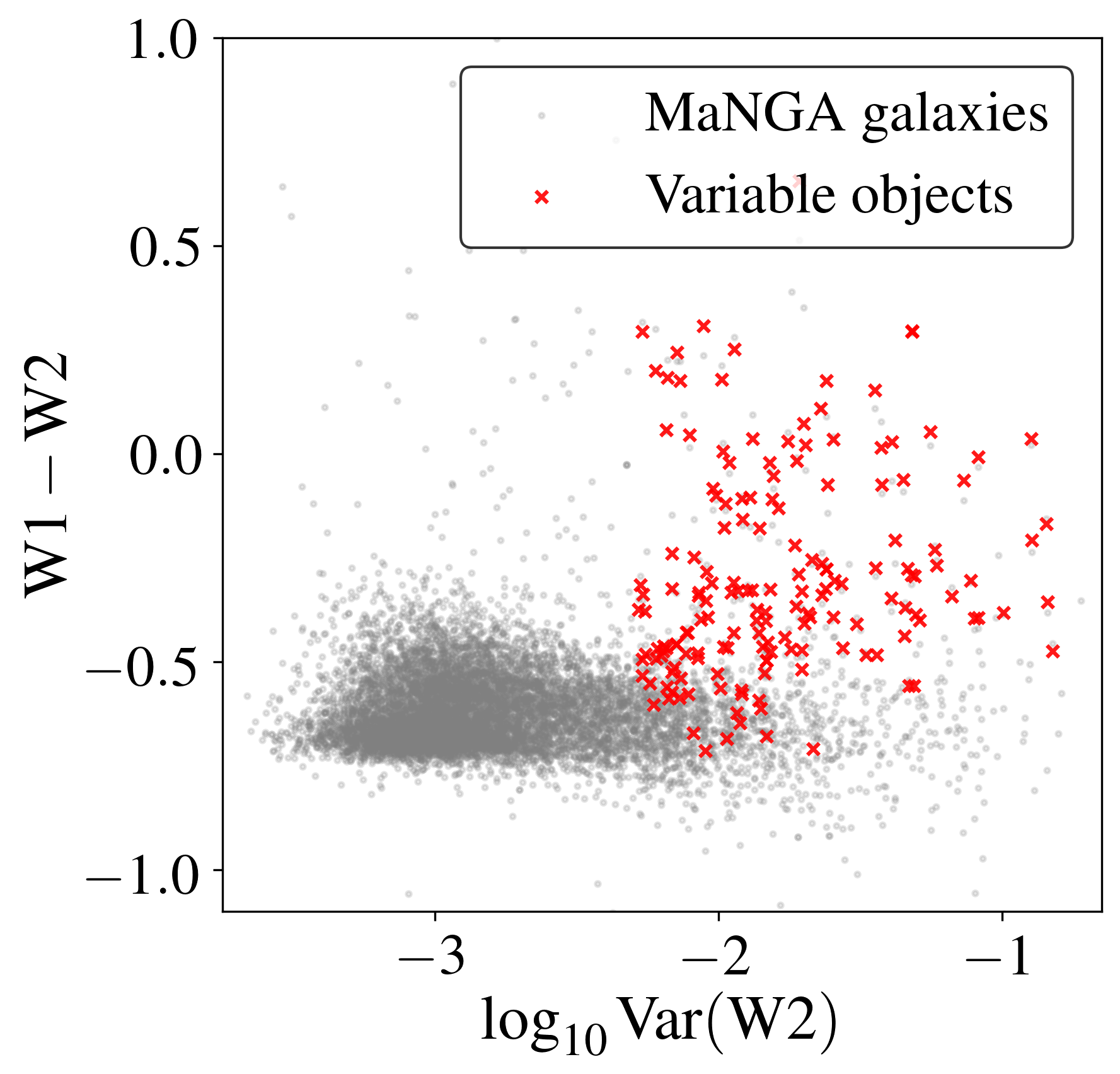}
    \caption{$W1-W2$ (AB) colors vs. $\log_{10}\Var{(W2)}$ of our sample. The red points represent our mid-IR variable objects.}
    \label{fig:varw2_w12}
\end{figure}

Figure \ref{fig:varw2_w12} shows the $W1-W2$ color of our sample 
versus $\log_{10}\Var{(W2)}$, providing an insight into the color 
distribution of galaxies as a function of their $W2$ variability. 
Our selection criteria allow objects with high $W2$ variability 
to be selected irrespective of their non-AGN-like colors. Nevertheless,
the identified variable objects tend to have red colors, which
indicates that they are mid-IR AGN (Figure \ref{fig:spsmodels}).

\begin{figure}[h]
    \centering
    \includegraphics[width=\columnwidth]{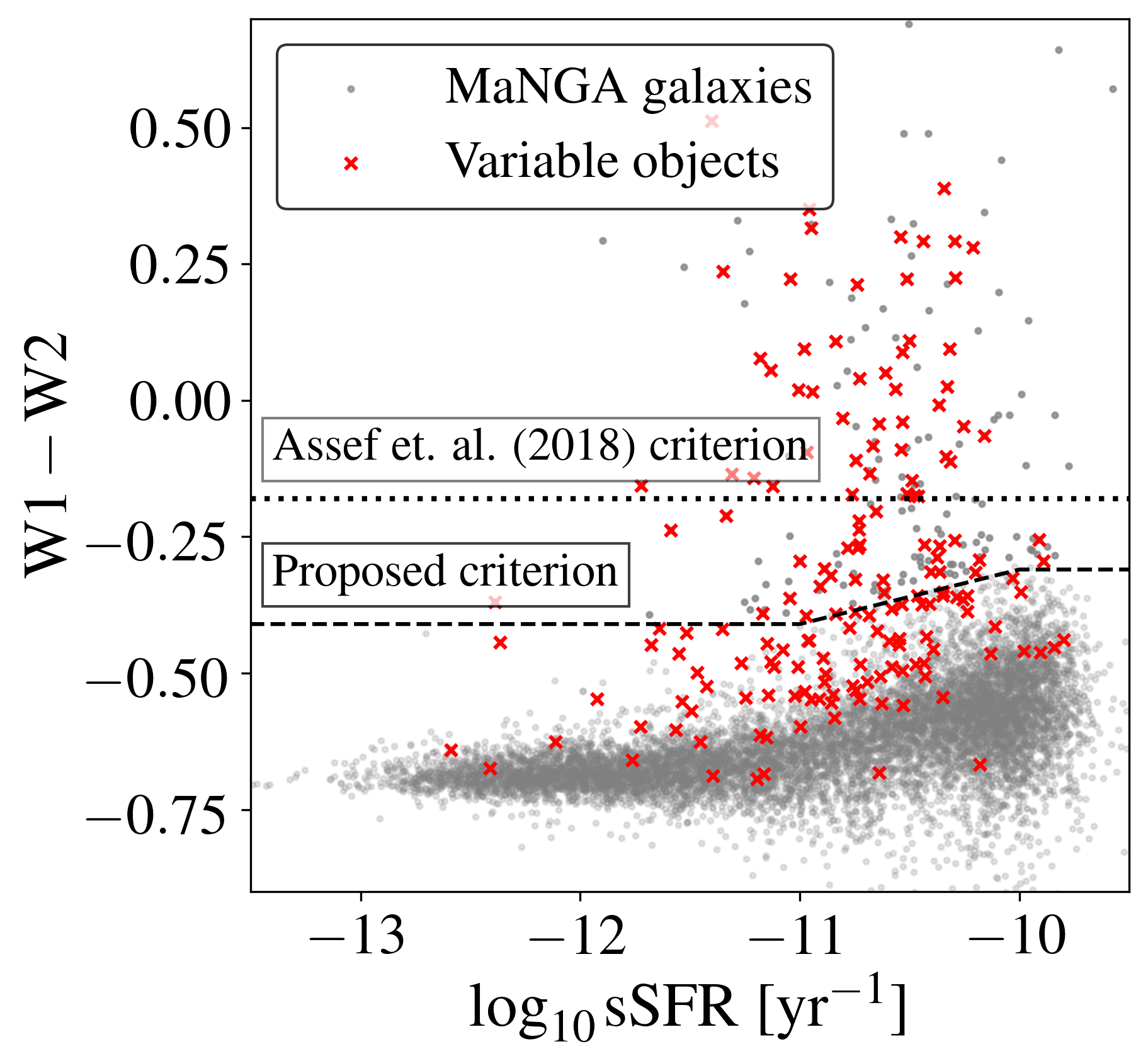}
    \caption{$W1-W2$ (AB) colors of MaNGA galaxies vs. their specific star formation rate determined from stellar population analysis. The 
    red points represent our mid-IR variable objects.}
    \label{fig:ssfr_w12}
\end{figure}

Figure \ref{fig:ssfr_w12} shows the $W1-W2$ color of all the MaNGA 
galaxies as a function of their specific star formation rate 
($\rm sSFR$). the Pipe3D $\rm sSFR$ values are calculated from the MaNGA aperture. Most galaxies in this plot lie on a tight sequence with
blue colors typical of stellar populations, while AGN host galaxies 
are redder in  $W1-W2$, and therefore skewed upwards 
(cf. Fig.~\ref{fig:spsmodels}). 

Non-AGN galaxies lie well below the cutoff set by \citet{assef18a},
where we use their least conservative (bluest) cutoff, converted to 
AB magnitudes. Galaxies that we classify as variable are shown in red, 
and extend into the star-forming sequence below the cutoff.

We propose a modified color criterion that extends the $W1-W2$ cutoff 
closer  to the main sequence of star-forming galaxies. 
Many of the galaxies above this threshold (and some below) are
detectably variable, supporting the interpretation of galaxies with 
these colors as AGN. 
Overall, the majority of the galaxies selected using our color criteria 
($\sim 80\%$) have a match as a broad line, optical narrow line, 
mid-IR variable, or red $W1-W2$ (using the \citealt{assef18a} cut) AGN.
This criterion therefore allows for a more complete population of AGN in 
nearby galaxies from their colors, with strong evidence that most if 
not all of these galaxies host true AGN. 

Using the criterion from \citet{assef18a}, about 29\% of the galaxies 
that we classify as variable are mid-IR color selected AGN. Using 
our modified criterion, around 46\% of the variable galaxies are 
mid-IR color selected AGN.

\subsubsection{MIRONGs}
\citet{jiang21a} describes a catalog of Mid InfraRed Outbursts in Nearby Galaxies (MIRONGs), which they define as galaxies that display a flare of
at least 0.5 mag in the mid-IR preceded by a stable phase without variability. The catalog was created from a sample selected from galaxies in the SDSS DR14 (\citealt{abolfathi18a}) by looking at their WISE variability from the NEOWISE database. Although their parent sample has significant overlap with MaNGA, our criterion is much more sensitive which is why only a small fraction of our galaxies appear as Mid-IR outbursts, which are more similar to changing look AGN or TDEs. 
Around 2\% (4 galaxies) of our variable objects have matches to MIRONGs from this catalog.

\subsubsection{Tidal Disruption Events}

\citet{masterson24a} identifies 18 Tidal Disruption Events (TDEs) in nearby galaxies ($< \SI{200}{\mega\parsec}$, or $z <0.05$) that were found using NEOWISE data. They look for transients across the NEOWISE database, and then cross-match it with the Census of the Local Universe (CLU; \citealt{cook19a}), thus enforcing the \SI{200}{\mega\parsec} criteria. The population of galaxies that they end up with therefore may not necessarily have a large overlap with the MaNGA sample.

They select events to have light-curves with a stable period before a rapid flare, followed by a slow monotonic decline. They also ensure that the $W2$ light-curve luminosity is greater than $\SI{e42}{\erg\per\s}$ and that the galaxy has non-AGN WISE colors. We find one match between our mid-IR variable objects and these TDE candidates, which they denote
``WTP 16aatsnw'' and which corresponds to plate-IFU 11981-6104. 
They conclude that this galaxy
is an AGN based on its location in BPT diagrams made using SDSS 
archival spectroscopy. However, they note that this object does not 
have or X-ray emission. As we note below, its $W1-W2$ colors are variable,
indicative of hot dust during its bright period and consistent with a 
stellar population otherwise.

Extreme Coronal Line Emitting galaxies are TDEs whose spectra contain very high ionization forbidden spectral lines such as $\text{[Fe VII]}$, $\text{[Fe X]}$ and $\text{[Fe XIV]}$. \citet{wang12a} finds seven galaxies from the SDSS DR7 (\citealt{abazajian09a}) with $z<0.38$ whose spectra show the presence of these coronal lines. \citet{callow24a} adds seven new candidates to this catalog of ECLEs by searching through the SDSS Legacy DR17 (\citealt{abdurro’uf22a}). We look for matches between our variable galaxies and these 14 ECLEs but do not find any, as these galaxies are not present in the MaNGA sample.

\subsubsection{Comments on Properties of Mid-IR Variable Galaxies}

The overlap of our variable catalog with AGN samples clearly shows that many
of the identified variables are associated with AGN. AGN identified with broad lines, 
narrow lines, and mid-IR colors are clearly overrepresented in the variable sample.
We find that 74\% of our mid-IR variable galaxies show one of these additional AGN signatures, using the \citet{assef18a} criterion
for WISE color.  This number increases to 80\% if we use our WISE color criterion instead.
%that only 22\% of our mid-IR variable galaxies do not show any additional AGN signatures. This number reduces to 20\% using our WISE color criterion.

That said, we caution the reader against inferences about the true AGN population from
the numbers in Table \ref{tab:classification}. The overlap among these samples
is affected by selection effects in all of the AGN signifiers, and correct
inferences require a much more detailed analysis accounting for such effects. 
For this purpose, Figure \ref{fig:manga_var_magcut} demonstrates that we can 
can detect variability with $\log_{10} \Var{(W2)} > - 2.3$, for all galaxies
brighter than  $W2 \sim 16.5$. Even in 
this case, the relative $W2$ brightness of the AGN relative to the host galaxy 
needs to be accounted for, as a 7\% variation in total flux (as seen in a majority of our mid-IR variable objects) may reflect a much larger variation of the AGN flux. Future work is required to infer the true joint distribution
of these AGN manifestations.

\section{Examples of Mid-IR Variable Galaxies \label{sec:var_examples}}
\begin{figure*}[]
    \centering
    \includegraphics[width=\textwidth]{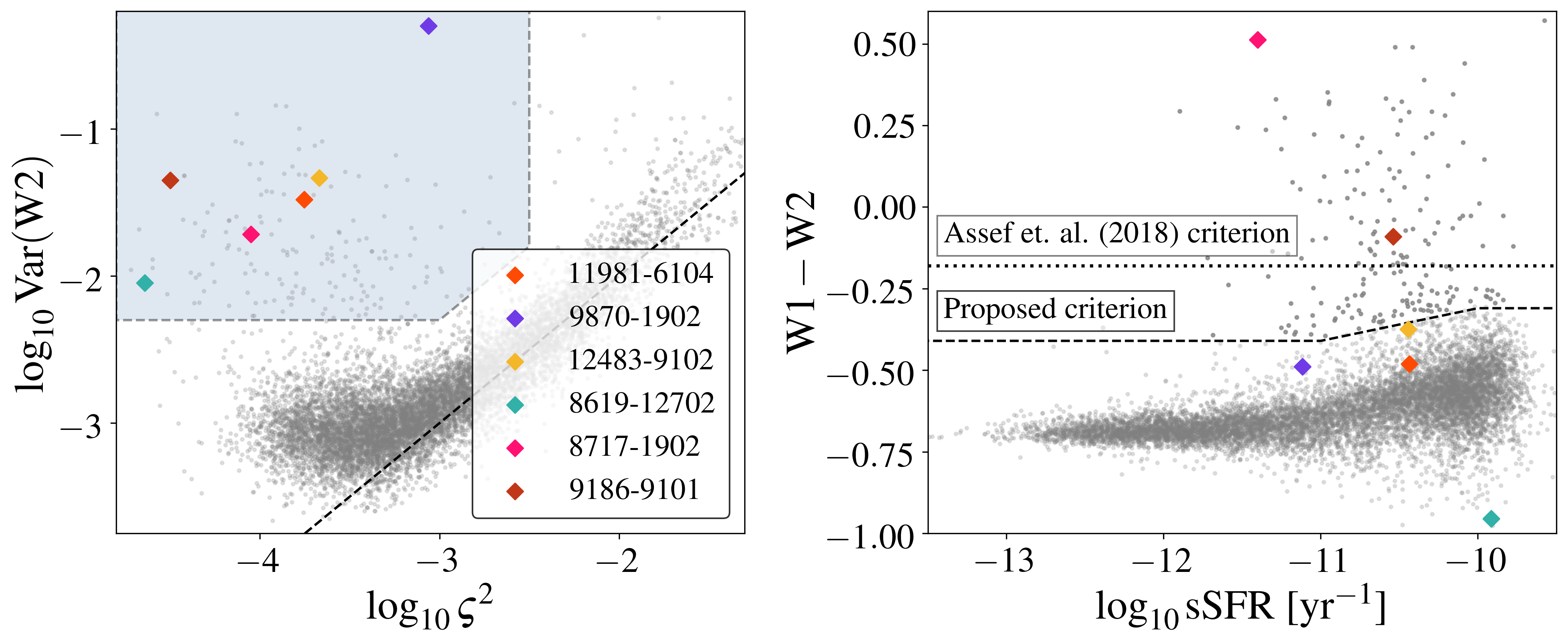}
    \caption{Showing where galaxies mentioned in Section~\ref{sec:var_examples} lie in our variability space (left panel) and in $W1-W2$ vs. sSFR space (right panel). Each galaxy is labelled by a diamond and labelled by its plate-IFU.}
    \label{fig:manga_image_objs}
\end{figure*}

% \begin{figure}[h!]
%     \centering
%     \includegraphics[width=\columnwidth]{manga_image_objs_vertical.png}
%     \caption{Location of example galaxies in Section~\ref{sec:var_examples} in our variability space (top panel) and in $W1-W2$ vs. sSFR space (bottom panel). Each galaxy is labelled by a diamond and labelled by its plate-IFU.}
%     \label{fig:manga_image_objs}
% \end{figure}

In this section, we present MaNGA central spectra for a small sample of
the mid-IR variable galaxies.
We also present optical broad band images of the galaxy synthesized
from the MaNGA data cube, and maps of the 
$\rm H\alpha$ and $\rm [OIII] 5007\AA$ emission lines, showing the fluxes of
those lines in the red and green channels, respectively. Finally, 
we plot the $W2$ light-curve of the galaxy to provide a comprehensive 
picture of the optical properties and mid-IR variability of the galaxy. 

These figures can be found in Appendix~\ref{app:images}. Figure~\ref{fig:manga_image_objs} shows where these objects lie in our variability space as well as $W1-W2-{\rm sSFR}$ space.

\begin{itemize}
    
\item \textbf{11981-6104}:
See the images of this galaxy in Fig.~\ref{fig:11981-6104and9870-1902}. This object has matches in both the MIRONG catalog and the \citet{masterson24a} TDE catalog. \citet{masterson24a} determines that it is a narrow line AGN (which
we also find by our criteria) and for that reason rules it out as a TDE. In our stacked WISE images it has mid-IR colors 
consistent with stellar, but during the increase in brightness it becomes
much redder (up to $W1-W2 \sim 0.5$), consistent with hot dust emission.

\item \textbf{9870-1902}:
See the images of this galaxy in Fig.~\ref{fig:11981-6104and9870-1902}. This object has the highest observed variability of all our mid-IR variable galaxies. This galaxy does not appear in any other AGN classification scheme (it has LINER-like narrow line ratios). Like the previous example 11981-6104,
it has stellar-like mid-IR colors before 2018, but during its bright phase
has much redder colors indicative of hot dust.

\item \textbf{12483-9102}:
See the images of this galaxy in Fig.~\ref{fig:12483-9102and8619-12702}. This object is the dwarf galaxy NGC 4395, which is a well-known low luminosity AGN host. It is red in $W1-W2$ colors (using our criterion) and is classified as a narrow-line AGN. 

\item \textbf{8619-12702}:
See the images of this galaxy in Fig.~\ref{fig:12483-9102and8619-12702}. This object appears in the broad-line AGN catalog.

\item \textbf{8717-1902:}
See the images of this galaxy in Fig.~\ref{fig:8717-1902and9186-9101}. This object has one of the reddest $W1-W2$ colors in our sample. It does not manifest as an AGN in broad or narrow line classifications, though it is 
near the boundary of the narrow line AGN criterion in Figure \ref{fig:p13},
with $P1\approx 0.10$ and $P3 \approx 0.47$.

\item \textbf{9186-9101:}
See the images of this galaxy in Fig.~\ref{fig:8717-1902and9186-9101}. This object is red in $W1-W2$ colors and appears in the broad-line catalog.

\end{itemize}

\section{\label{sec:summary}Summary}

We created a mid-IR variability catalog of galaxies, using the MaNGA component of SDSS-IV and their $W2$ light-curves from the NEOWISE and WISE All-Sky surveys. 
We used the observed variance across all epochs and compared it to the expected variance in $W2$ to identify variable objects, using the within-epoch
variances in $W2$ to provide more accurate across-epoch expected 
variances. We used a Monte Carlo simulation to determine robust criteria 
for variability. 
We find that 170 galaxies satisfy these variability criteria.
We also show that our catalog is roughly
complete for variability with $\log_{10} \Var{(W2)} > -2.3$ for 
galaxies with $W2<16.5$. 
The galaxies are listed in Appendix~\ref{app:catalog}. 

We compare our catalog to existing broad-line and narrow-line AGN catalogs, 
MIRONG and TDE catalogs, and use traditional mid-IR AGN classification schemes such as the $W1-W2$ color to illuminate the demographics of the population of variable objects that we have selected. The majority 
of the variable objects ($\sim 80\%$) have other signatures of AGN
activity.

As part of this work, we propose a modified $W1-W2$ criterion for the 
MaNGA sample that provides a more complete sample of the population
of AGN (Section \ref{sec:w12}).

\section{\label{sec:dataavail}Data Availability}

The data used in this article (particularly in Fig.~\ref{fig:var_objs}) have been tabulated in Appendix \ref{app:catalog}. The $W2$ light-curves of all mid-IR variable galaxies have been plotted in Appendix \ref{app:light-curves}. The full MANGA-WISE variability catalog is available in the online supplementary material, which contains all of the statistical quantities calculated in Section~\ref{sec:catcontents} for all galaxies in MaNGA's DR17. A key between the names used in this article and the names in the file is also provided in Appendix~\ref{app:catalog}.

\section{Acknowledgements}
%\begin{acknowledgements}
We thank David G.~ Grier, David W.~Hogg, Or Graur, Sjoert Van Velzen and the IRSA helpdesk
for useful conversations and guidance in writing this paper. 
AP acknowledges the support of the Dean's Undergraduate Research Fund
from the College of Arts \& Sciences at New York University.

This research has made use of the NASA/IPAC Infrared Science Archive, which is funded by the National Aeronautics and Space Administration and operated by the California Institute of Technology.

This publication makes use of data products from the Wide-field Infrared Survey Explorer, which is a joint project of the University of California, Los Angeles, and the Jet Propulsion Laboratory/California Institute of Technology, funded by the National Aeronautics and Space Administration.

This publication also makes use of data products from NEOWISE, which is a project of the Jet Propulsion Laboratory/California Institute of Technology, funded by the Planetary Science Division of the National Aeronautics and Space Administration.

Funding for the Sloan Digital Sky 
Survey IV has been provided by the 
Alfred P. Sloan Foundation, the U.S. 
Department of Energy Office of 
Science, and the Participating 
Institutions. 

SDSS-IV acknowledges support and 
resources from the Center for High 
Performance Computing  at the 
University of Utah. The SDSS 
website is \url{www.sdss4.org}.

SDSS-IV is managed by the 
Astrophysical Research Consortium 
for the Participating Institutions 
of the SDSS Collaboration including 
the Brazilian Participation Group, 
the Carnegie Institution for Science, 
Carnegie Mellon University, Center for 
Astrophysics | Harvard \& 
Smithsonian, the Chilean Participation 
Group, the French Participation Group, 
Instituto de Astrof\'isica de 
Canarias, The Johns Hopkins 
University, Kavli Institute for the 
Physics and Mathematics of the 
Universe (IPMU) / University of 
Tokyo, the Korean Participation Group, 
Lawrence Berkeley National Laboratory, 
Leibniz Institut f\"ur Astrophysik 
Potsdam (AIP),  Max-Planck-Institut 
f\"ur Astronomie (MPIA Heidelberg), 
Max-Planck-Institut f\"ur 
Astrophysik (MPA Garching), 
Max-Planck-Institut f\"ur 
Extraterrestrische Physik (MPE), 
National Astronomical Observatories of 
China, New Mexico State University, 
New York University, University of 
Notre Dame, Observat\'ario 
Nacional / MCTI, The Ohio State 
University, Pennsylvania State 
University, Shanghai 
Astronomical Observatory, United 
Kingdom Participation Group, 
Universidad Nacional Aut\'onoma 
de M\'exico, University of Arizona, 
University of Colorado Boulder, 
University of Oxford, University of 
Portsmouth, University of Utah, 
University of Virginia, University 
of Washington, University of 
Wisconsin, Vanderbilt University, 
and Yale University.

\facility{IRSA, WISE}
%\end{acknowledgements}

% \begin{figure*}[h!]
%     \begin{center}
%         \includegraphics[width=12cm]{monte_carlo_dist_30percentvar_4x4.png}
%         \caption{\textbf{(a)} compares $\log_{10} \varsigma^2$ vs. $W2$, with the black line plotting the vertical edge of the variable cut in \textbf{(b)}. \textbf{(b)} show the population with $W2<16.5$. It is also labelled with 'count', which is the number of variable objects that remain after the magnitude cut.}
%         \label{fig:montecarlo4x4}
%     \end{center}
% \end{figure*}

\clearpage
\bibliography{references}{}
\bibliographystyle{aasjournal}

\appendix

\section{\label{app:images} Variable galaxy images}

\begin{center}
Here, we provide images, spectra and $W2$ light-curves of the galaxies mentioned in Section~\ref{sec:var_examples}. The images and spectra were obtained using the SDSS Marvin tool (\citealt{cherinka19a}). 
\end{center}

\begin{figure*}[h!]
    \centering
    \includegraphics[width=0.93\textwidth]{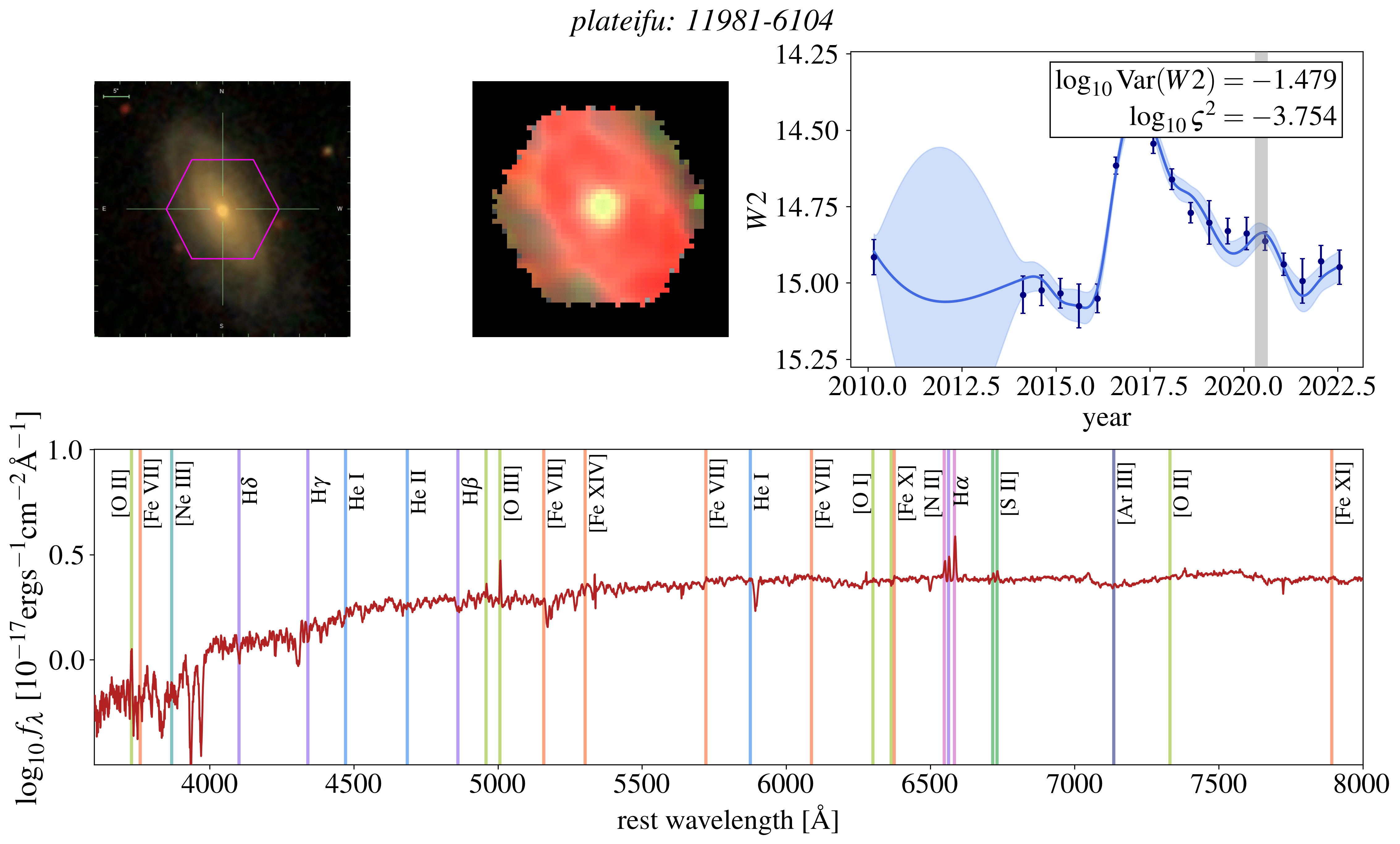}
\\
    \centering
    \includegraphics[width=0.95\textwidth]{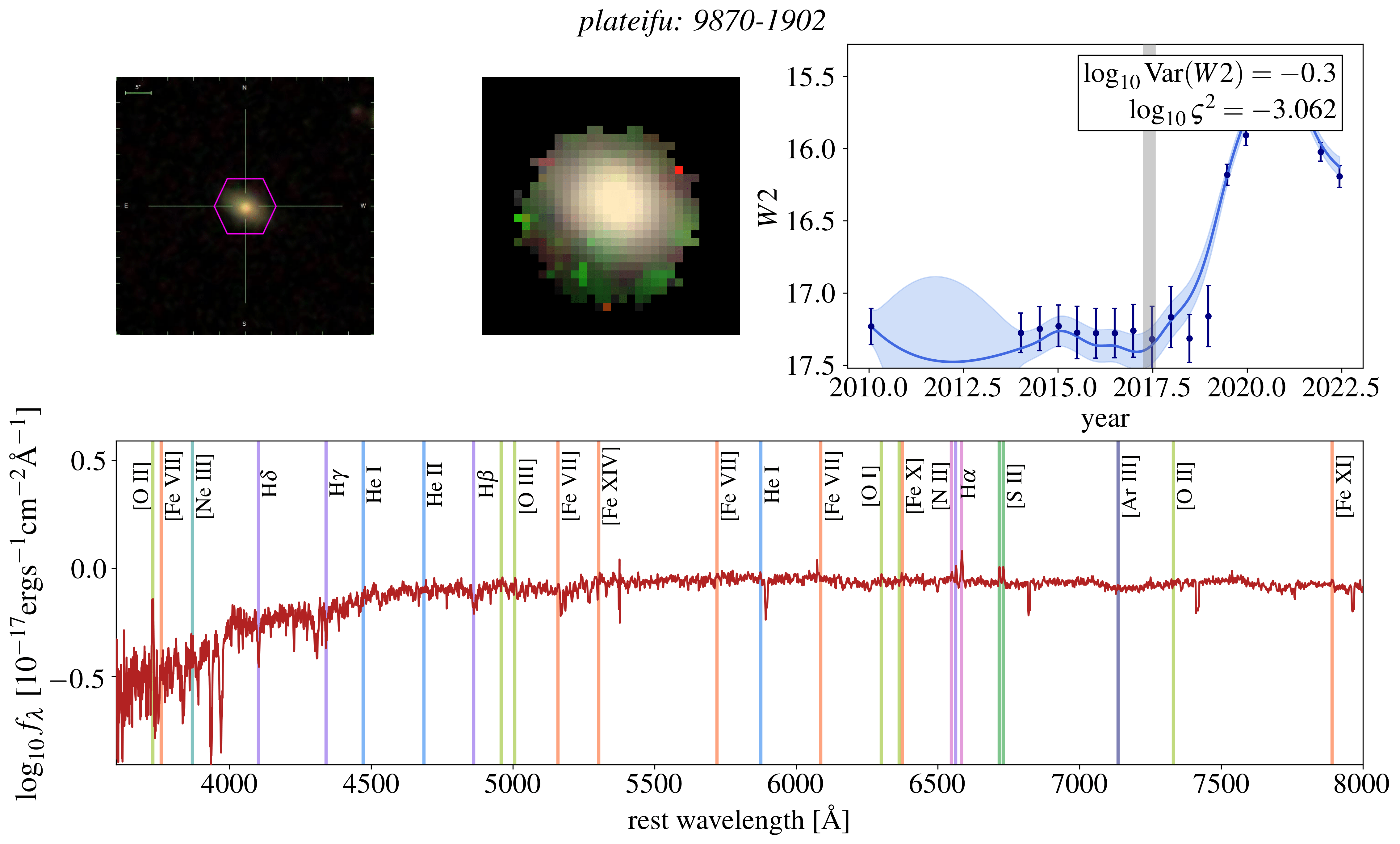}
    \caption{Images and spectra of galaxies 11981-6104 and 9870-1902. For each galaxy, the top left panel shows the SDSS optical image of the galaxy, the top center panel shows the H$\alpha$ (red) and $\rm [OIII]$ (green) maps overlaid on the $r$-band flux (grey). The top right panel shows the $W2$ light-curve (the grey line indicates when it was observed by MaNGA). The bottom panel shows the MaNGA spectra with emission lines marked. Lines from the same element are of the same color.}
    \label{fig:11981-6104and9870-1902}
\end{figure*}

\begin{figure*}[h!]
    \centering
    \includegraphics[width=\textwidth]{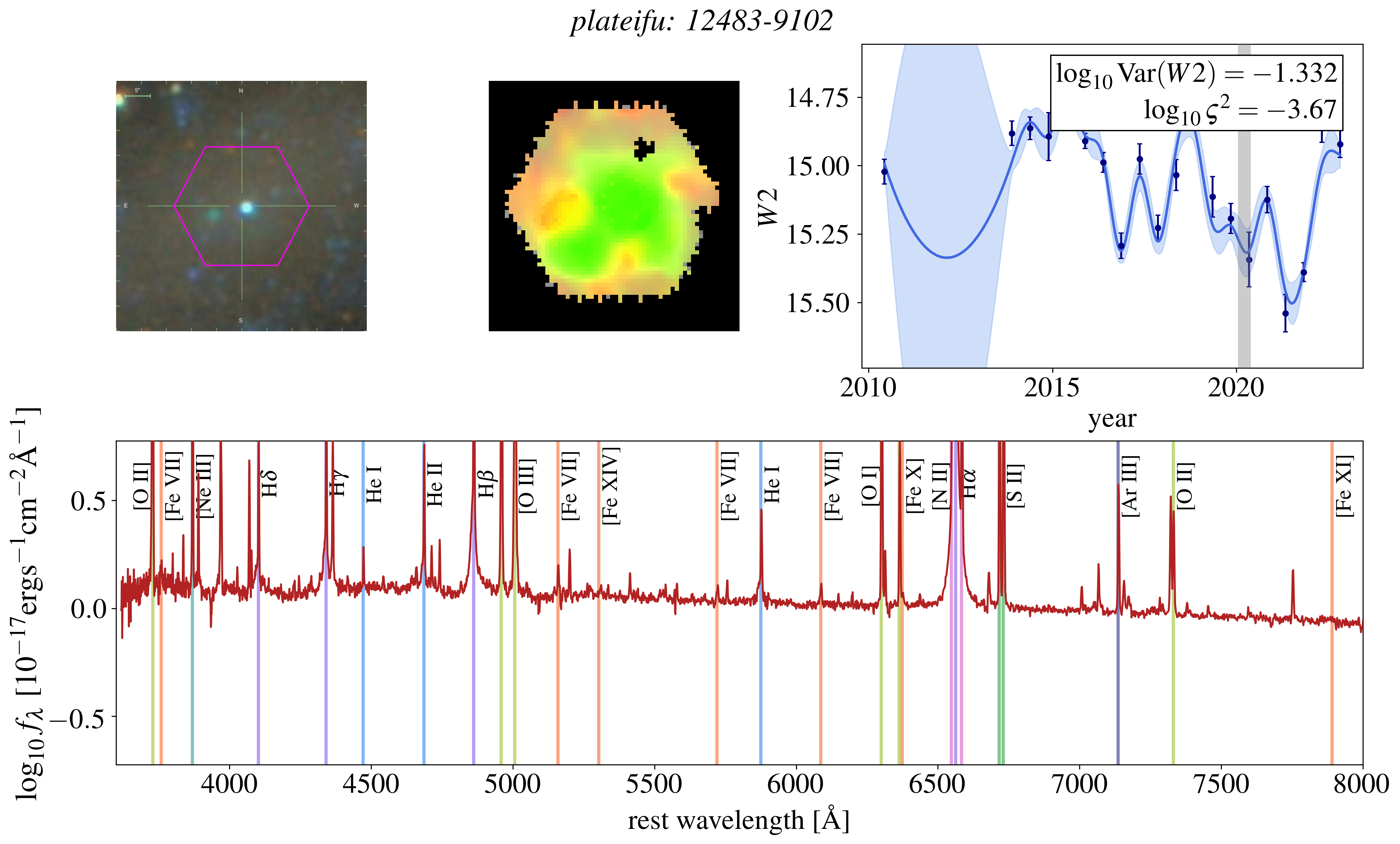}
\\
    \centering
    \includegraphics[width=\textwidth]{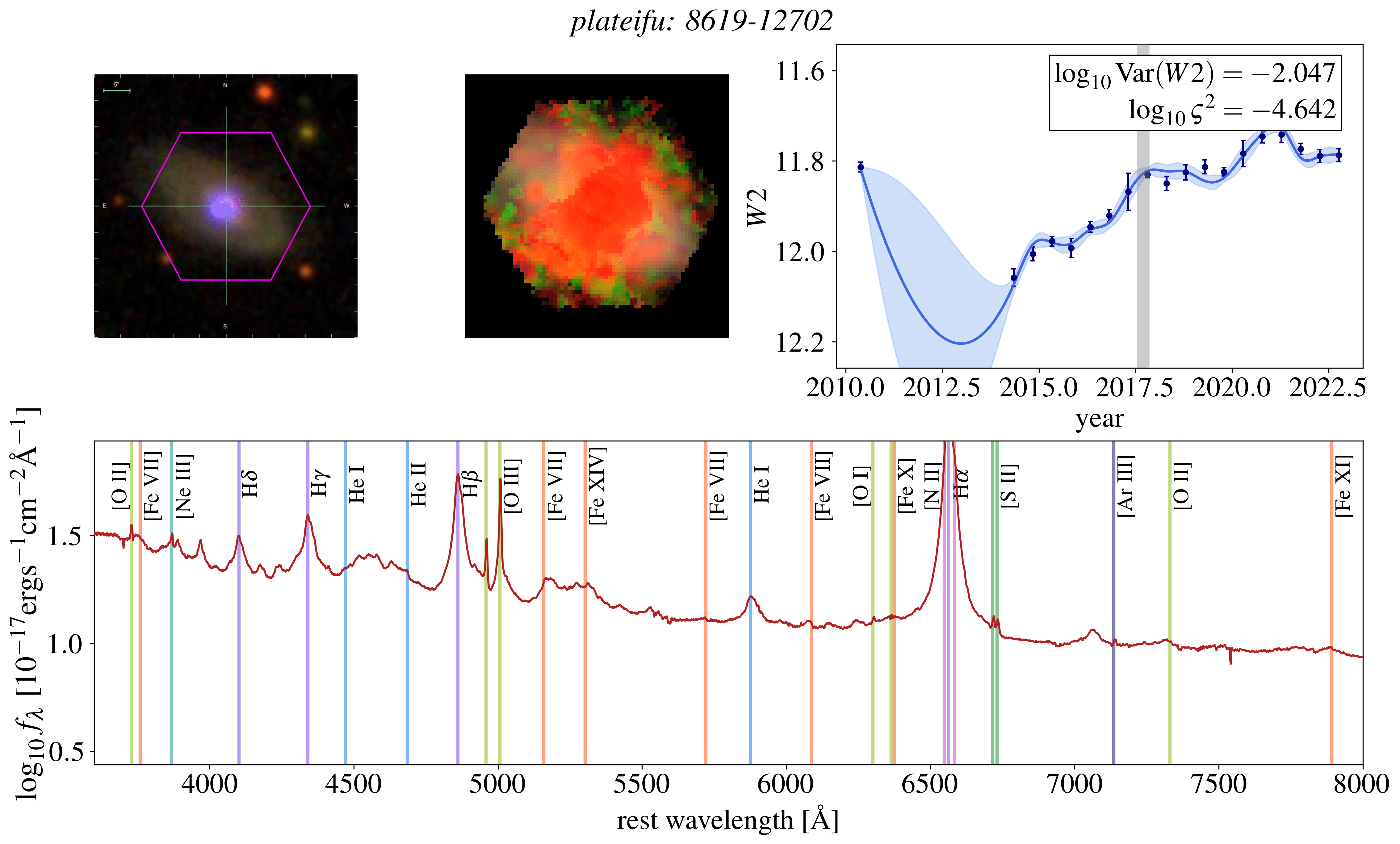}
    \caption{Similar to Figure \ref{fig:11981-6104and9870-1902}, for plate-IFUs 12483-9102 and 8619-12702. Plate-IFU 12483-9102 observed the center of NGC 4395, which is a well-known low mass AGN hosting dwarf galaxy. The image and MaNGA data cube only cover the central region of the whole galaxy, which subtends a large solid angle on the sky. }
    \label{fig:12483-9102and8619-12702}
\end{figure*}

\begin{figure*}[h!]
    \centering
    \includegraphics[width=\textwidth]{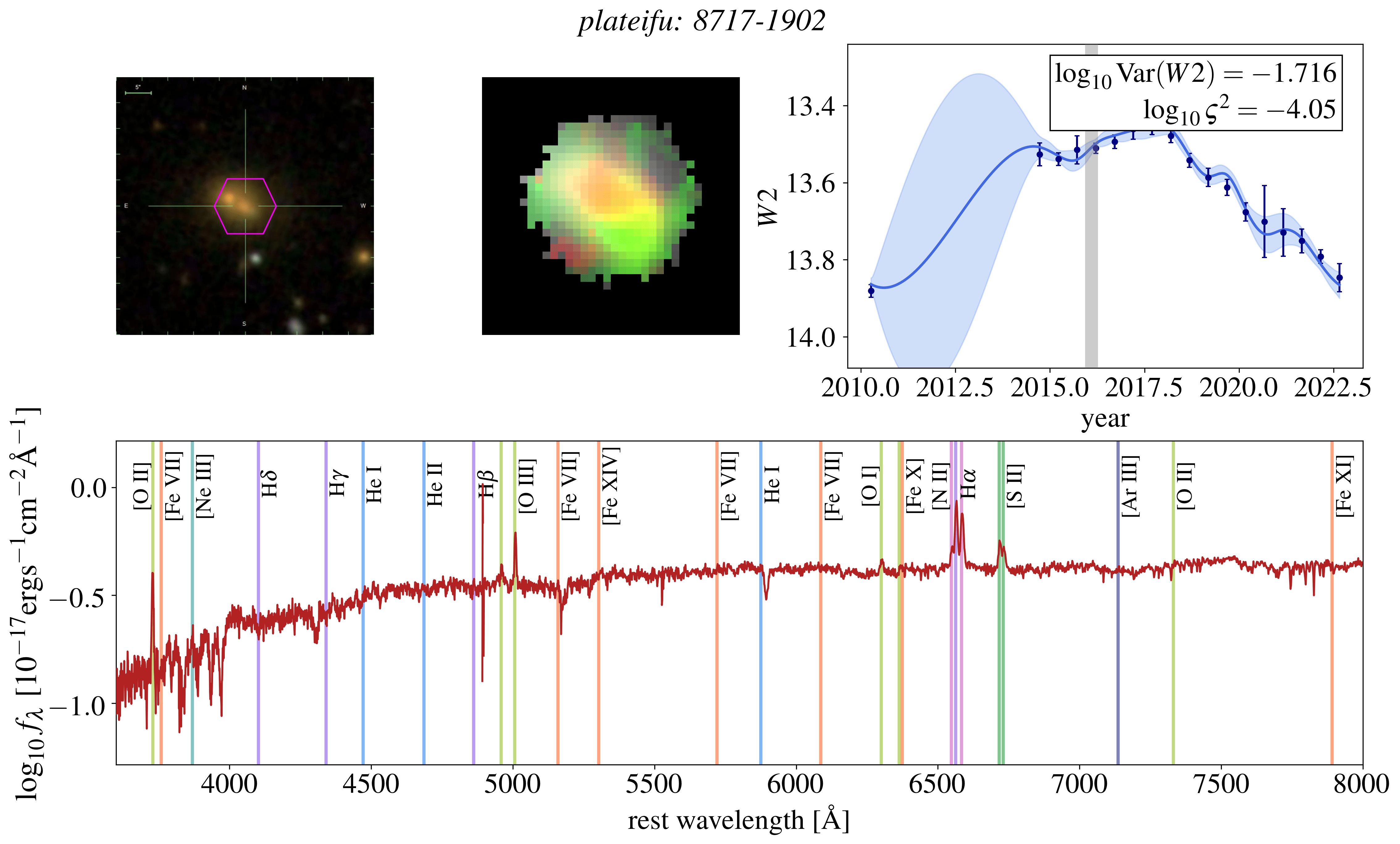}
\\
    \centering
    \includegraphics[width=\textwidth]{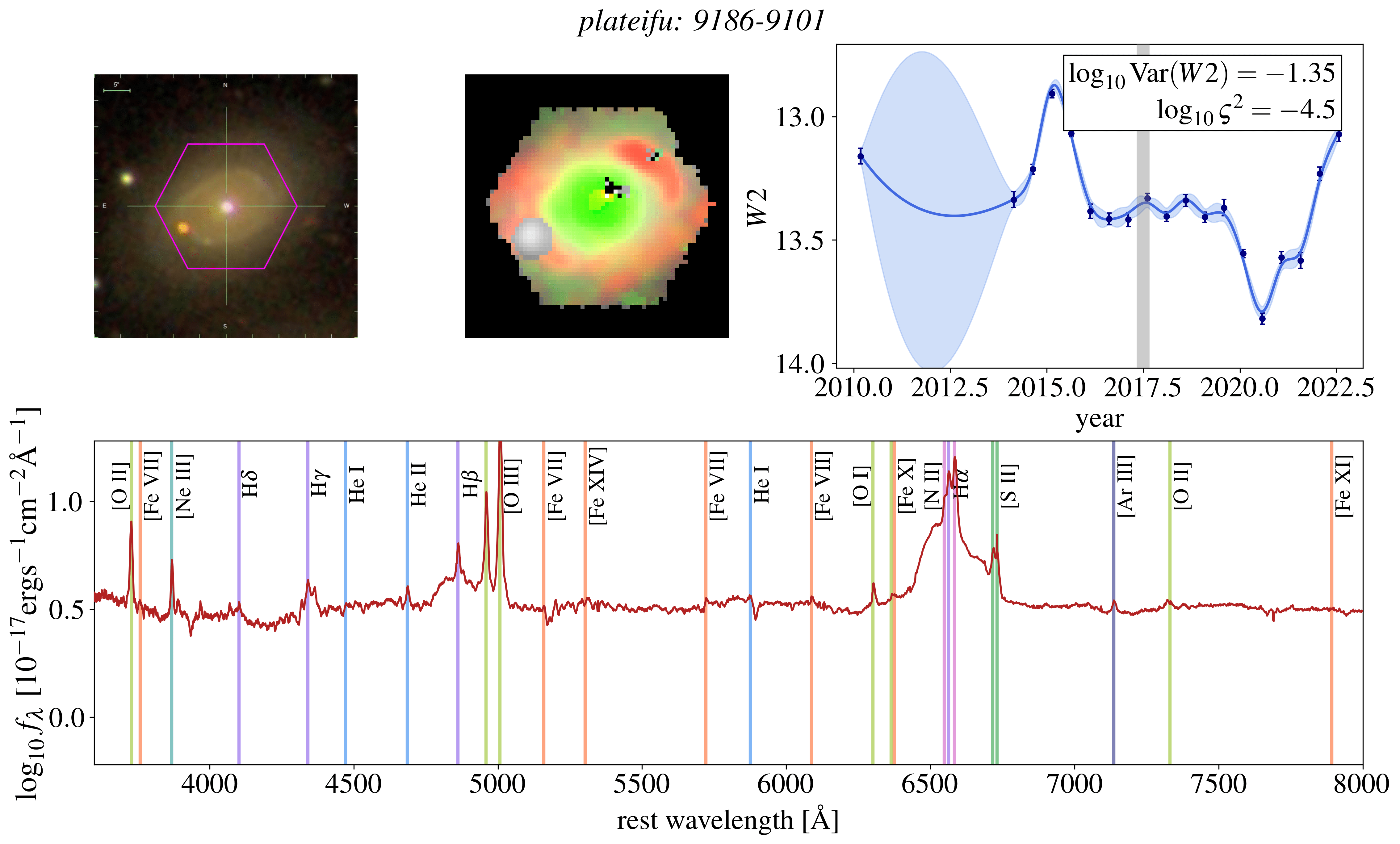}
    \caption{Similar to Figure \ref{fig:11981-6104and9870-1902}, for 
    plate-IFUs 8717-1902 and 9186-9101.}
    \label{fig:8717-1902and9186-9101}
\end{figure*}

\clearpage
\section{\label{app:catalog} The MaNGA-WISE Variability Catalog}

Table~\ref{tab:cat_defs} provides a key between the column labels in the variability catalog and the statistical quantities calculated in Section~\ref{sec:catcontents}. Here, `$W?$' can be either $W1$ or $W2$.
Table ~\ref{tab:cat1} lists all of the objects that we identify as variable in the mid-IR. 

\input{cat_defs}
\input{variability_catalog}

\clearpage
\section{\label{app:light-curves} Mid-IR Variable Galaxy light-curves}

We plot the $W2$ light-curves ordered by decreasing $\log_{10}\Var{(W2)}$. 
To interpolate between observations we perform Gaussian process regression with the {\tt scikit-learn} Python package, using
a Matern-3/2 kernel with a length scale hyperparameter of $\sim 15$ years. Light-curves of galaxies which have the {\tt vi\_flag} set are plotted in red. The gray bands mark the dates of the MaNGA observations.

 \begin{figure*}[h!]
 \centering
        \caption{$W2$ light-curves of mid-IR variable galaxies, ordered by $\log_{10}\Var{(W2)}$ and labelled by MaNGA plate-IFU (top right). They are also labelled by their coordinates on our variability space (top left), where ${\rm LV}(W2) \equiv \log_{10}\Var{(W2)}$ and ${\rm L}\varsigma^2 \equiv \log_{10}\varsigma^2$. The grey line marks when MaNGA observed the galaxy. The blue line and band show the Gaussian process regression fit. Galaxies with the {\tt vi\_flag} set are plotted in red.}
        \includegraphics[width=0.9\textwidth] {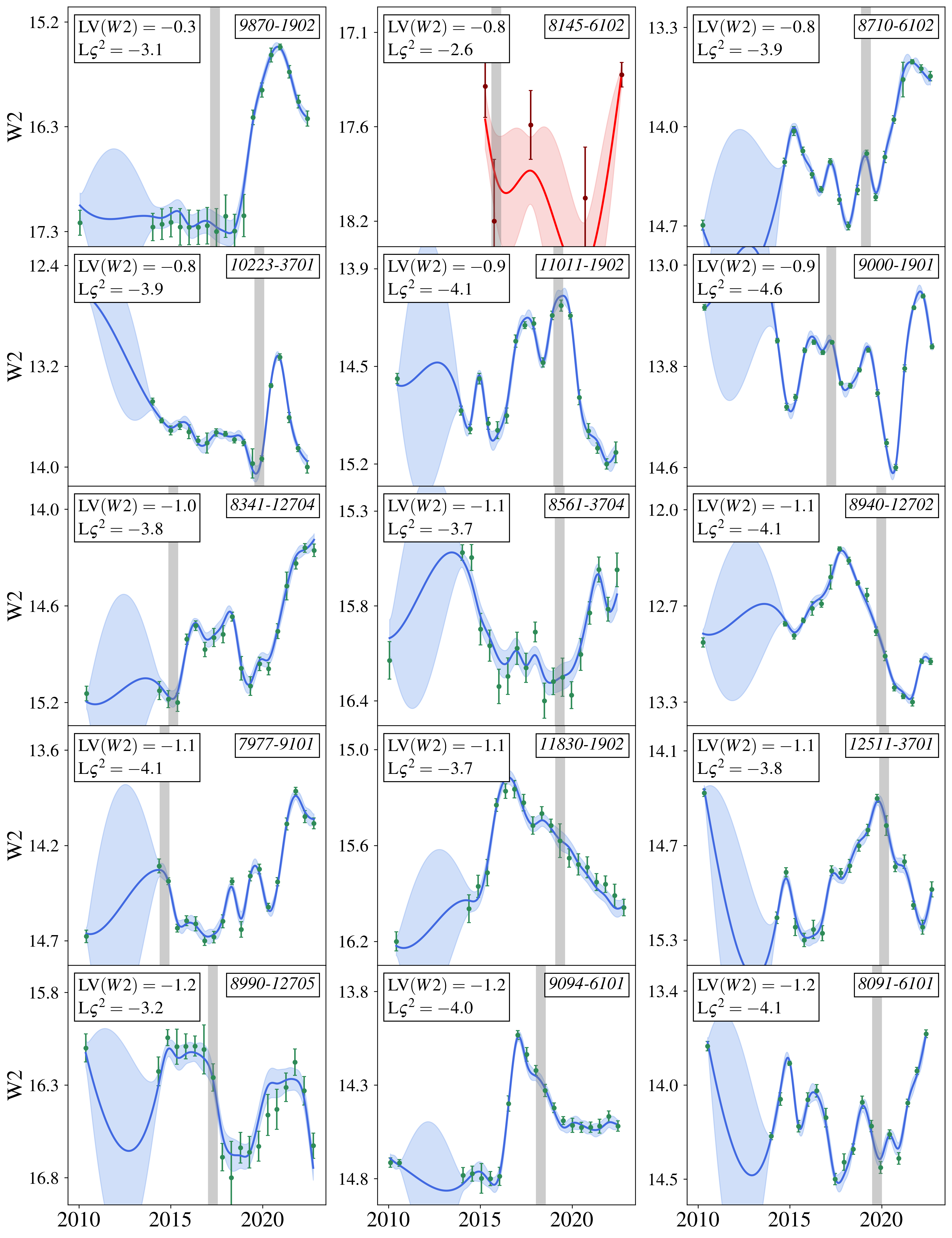}
     \label{fig:light-curves0}
  
 \end{figure*}

 \begin{figure*}[b!]
     \includegraphics[width=\textwidth]{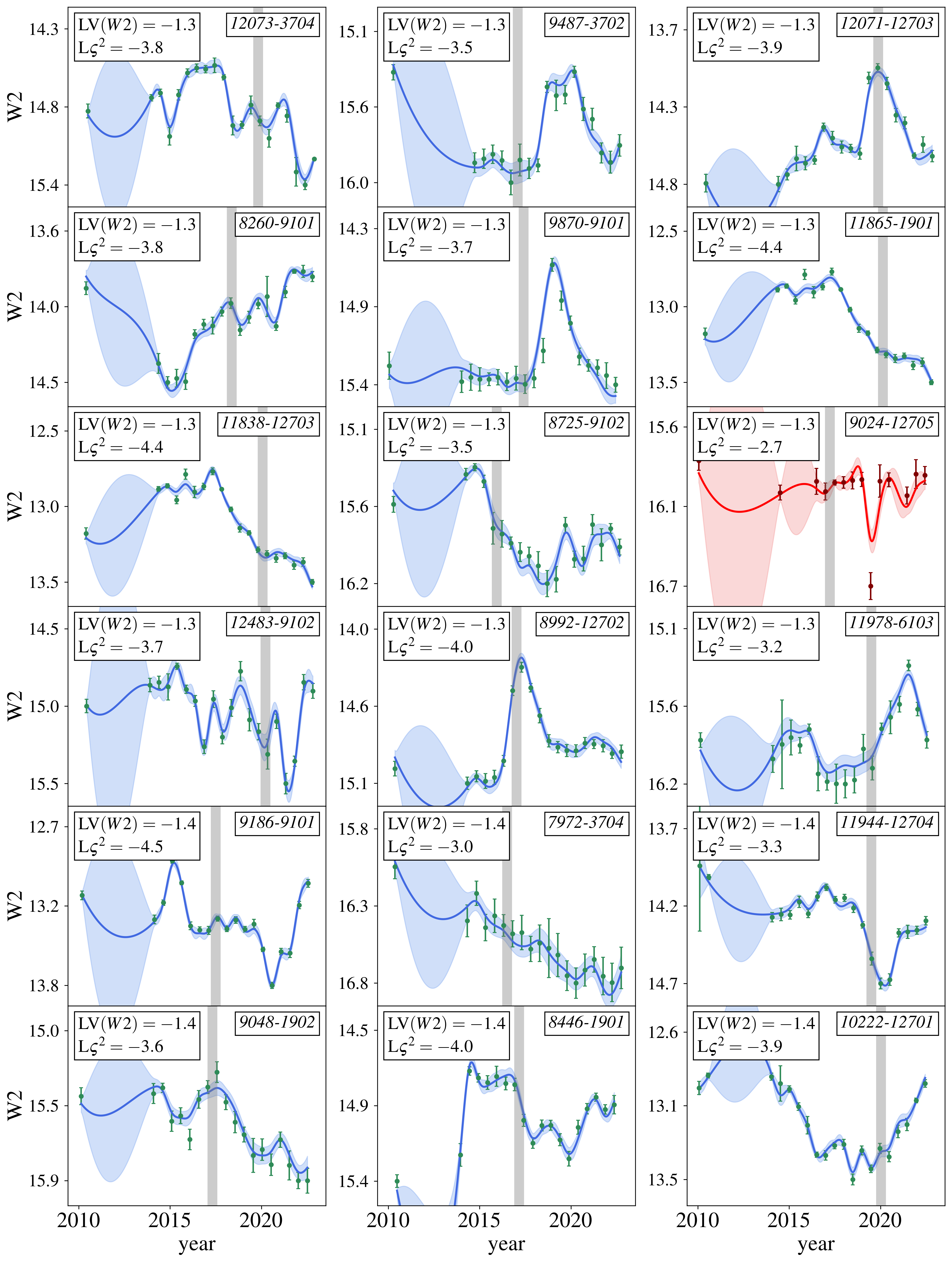}
     \label{fig:lightcurves10}
     %\caption{$W2$ light-curves of mid-IR variable galaxies, ordered by $\log_{10}\Var{(W2)}$ and labelled by MaNGA plate-IFU. The grey line marks when MaNGA observed the galaxy.}
 \end{figure*}

 \begin{figure*}[b!]
     \includegraphics[width=\textwidth]{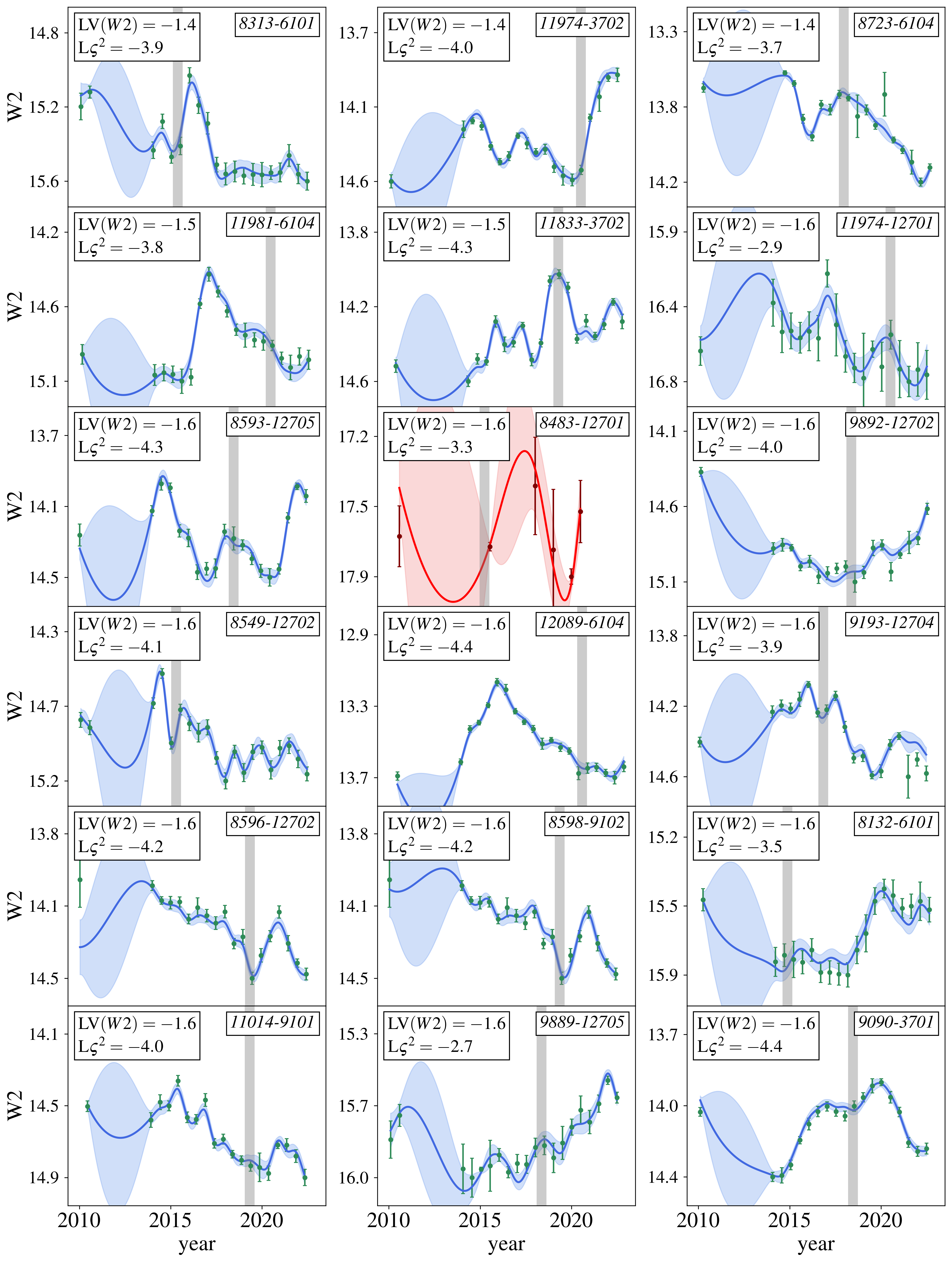}
     \label{fig:lightcurves20}
     %\caption{$W2$ light-curves of mid-IR variable galaxies, ordered by $\log_{10}\Var{(W2)}$ and labelled by MaNGA plate-IFU. The grey line marks when MaNGA observed the galaxy.}
 \end{figure*}

 \begin{figure*}[b!]
     \includegraphics[width=\textwidth]{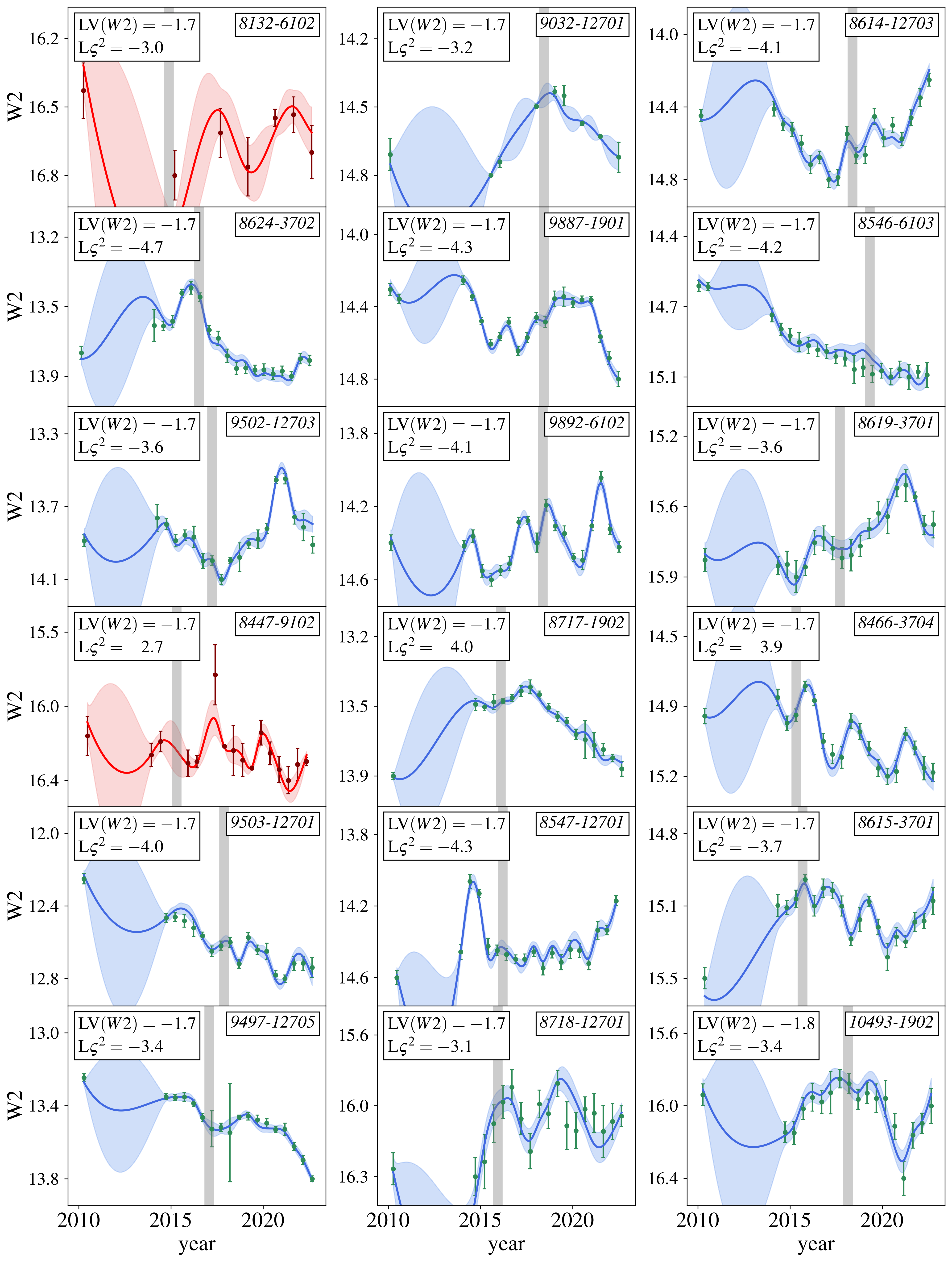}
    \label{fig:lightcurves30}
    %\caption{$W2$ light-curves of mid-IR variable galaxies, ordered by $\log_{10}\Var{(W2)}$ and labelled by MaNGA plate-IFU. The grey line marks when MaNGA observed the galaxy.}
 \end{figure*}

 \begin{figure*}[b!]
     \includegraphics[width=\textwidth]{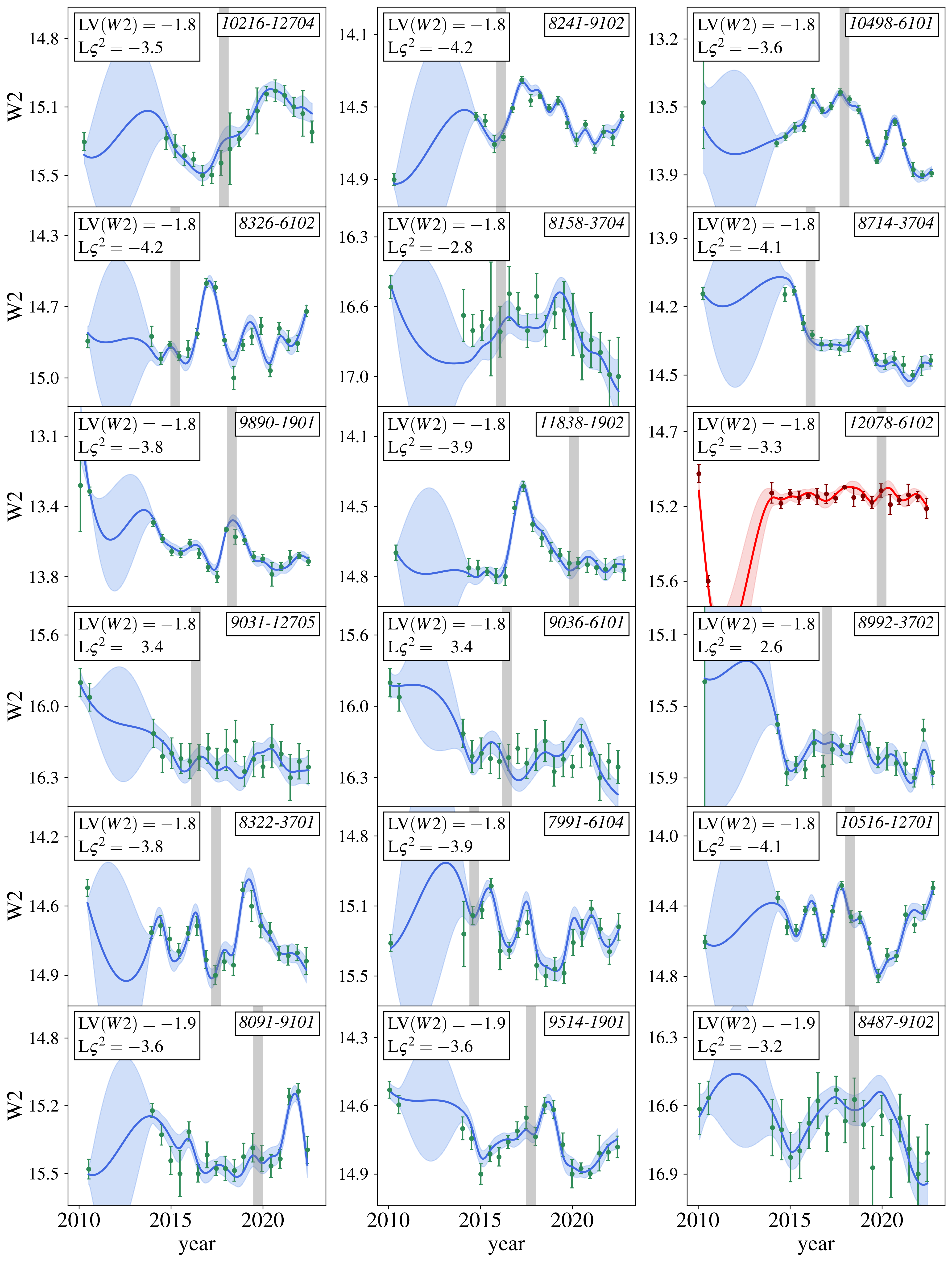}
     \label{fig:lightcurves40}
     %\caption{$W2$ light-curves of mid-IR variable galaxies, ordered by $\log_{10}\Var{(W2)}$ and labelled by MaNGA plate-IFU. The grey line marks when MaNGA observed the galaxy.}
 \end{figure*}

\begin{figure*}[b!]
     \includegraphics[width=\textwidth]{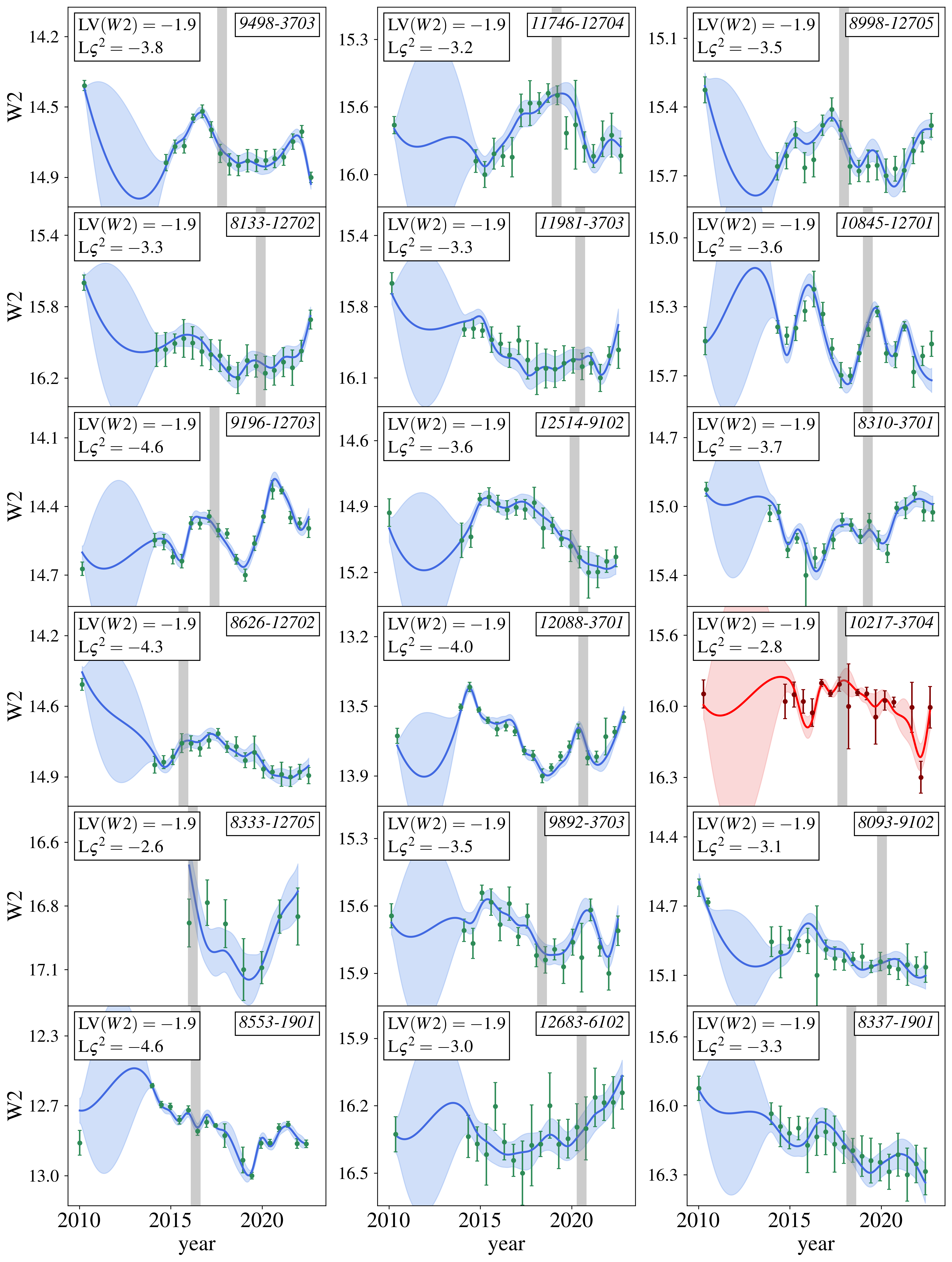}
     \label{fig:lightcurves50}
     %\caption{$W2$ light-curves of mid-IR variable galaxies, ordered by $\log_{10}\Var{(W2)}$ and labelled by MaNGA plate-IFU. The grey line marks when MaNGA observed the galaxy.}
 \end{figure*}

 \begin{figure*}[b!]
     \includegraphics[width=\textwidth]{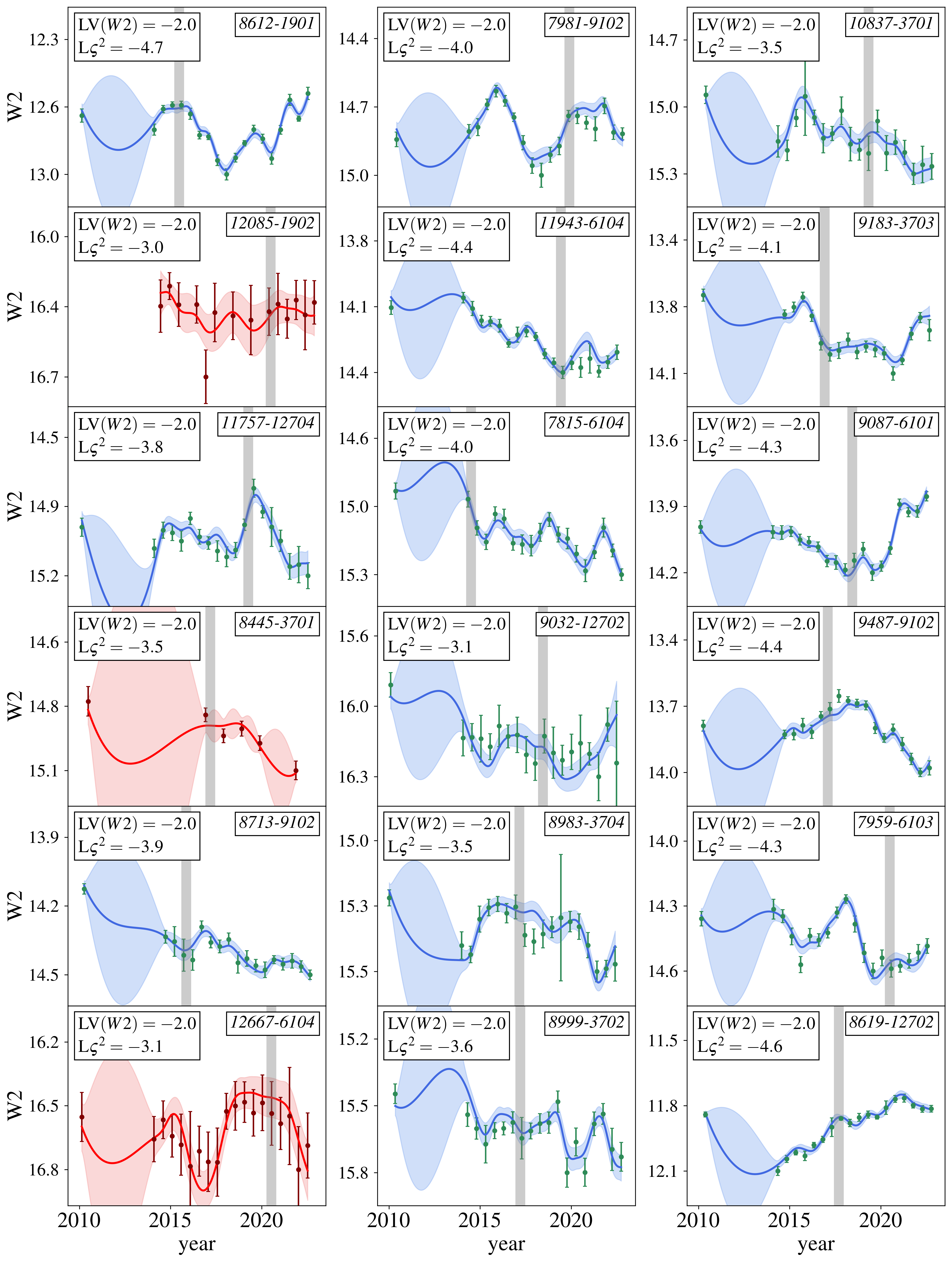}
     \label{fig:lightcurves60}
     %\caption{$W2$ light-curves of mid-IR variable galaxies, ordered by $\log_{10}\Var{(W2)}$ and labelled by MaNGA plate-IFU. The grey line marks when MaNGA observed the galaxy.}
 \end{figure*}

 \begin{figure*}[b!]
     \includegraphics[width=\textwidth]{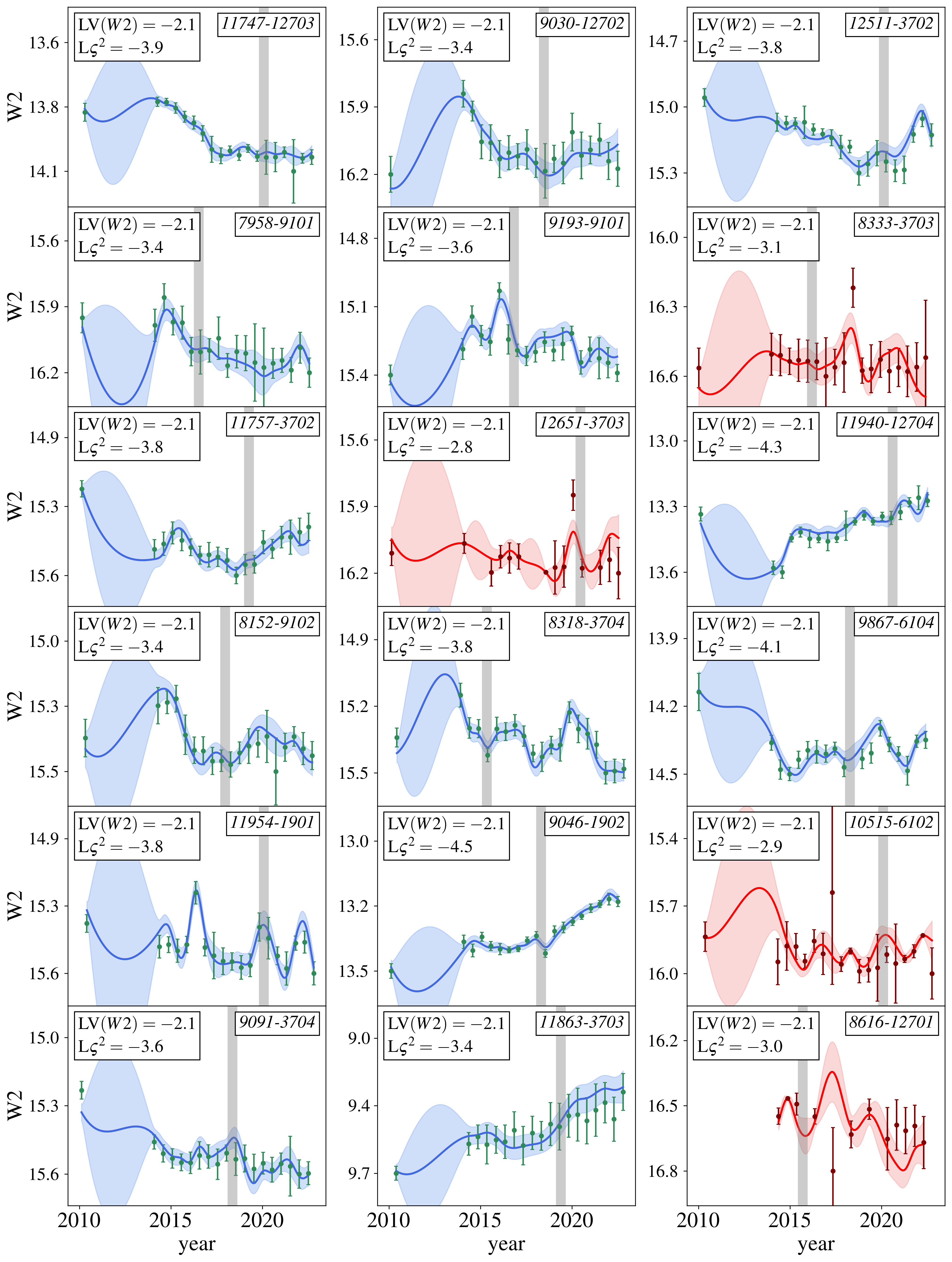}
    \label{fig:lightcurves70}
    %\caption{$W2$ light-curves of mid-IR variable galaxies, ordered by $\log_{10}\Var{(W2)}$ and labelled by MaNGA plate-IFU. The grey line marks when MaNGA observed the galaxy.}
 \end{figure*}

 \begin{figure*}[b!]
     \includegraphics[width=\textwidth]{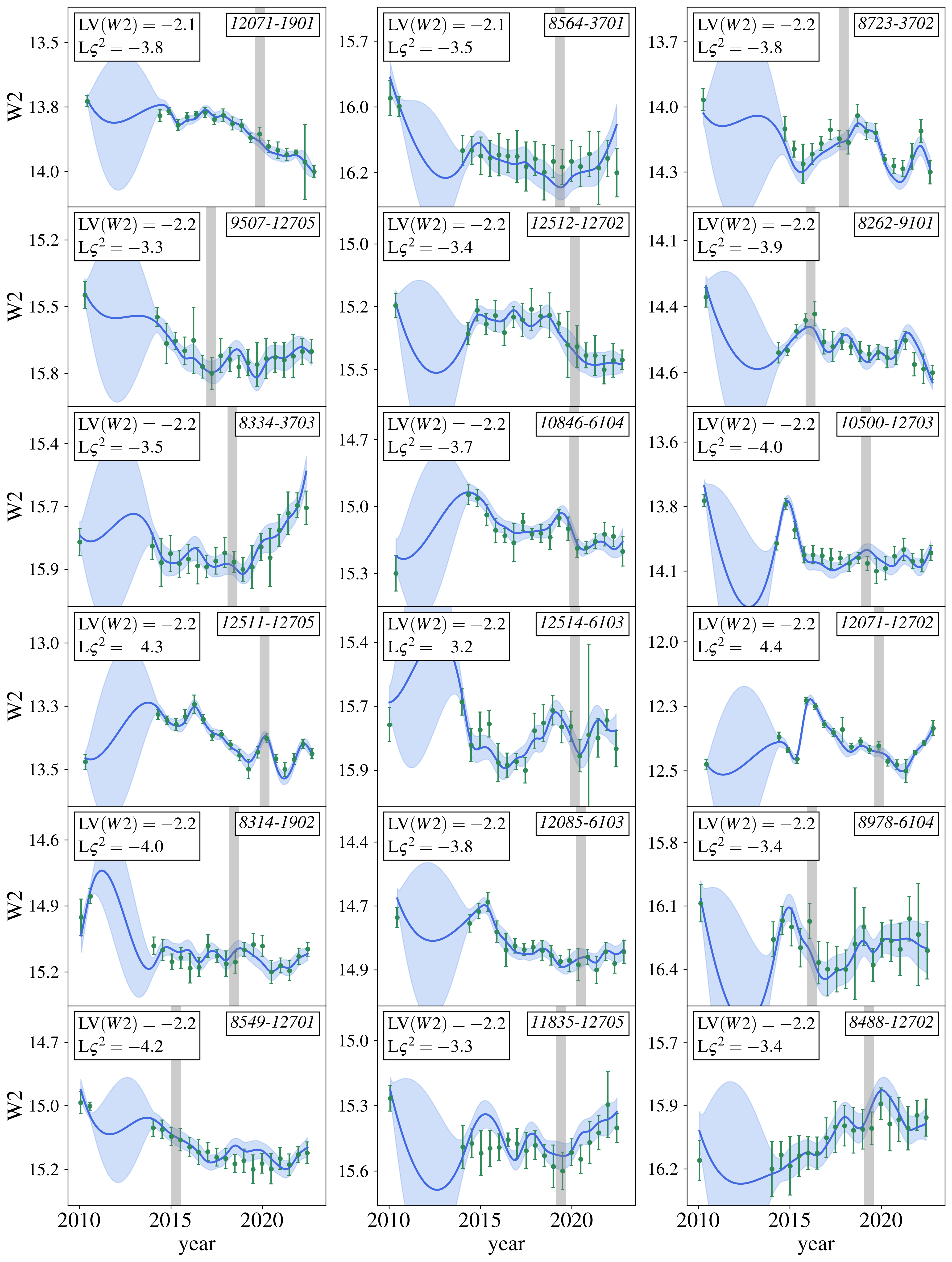}
     \label{fig:lightcurves80}
     %\caption{$W2$ light-curves of mid-IR variable galaxies, ordered by $\log_{10}\Var{(W2)}$ and labelled by MaNGA plate-IFU. The grey line marks when MaNGA observed the galaxy.}
 \end{figure*}
 
  \begin{figure*}[t!]
     \includegraphics[width=\textwidth]{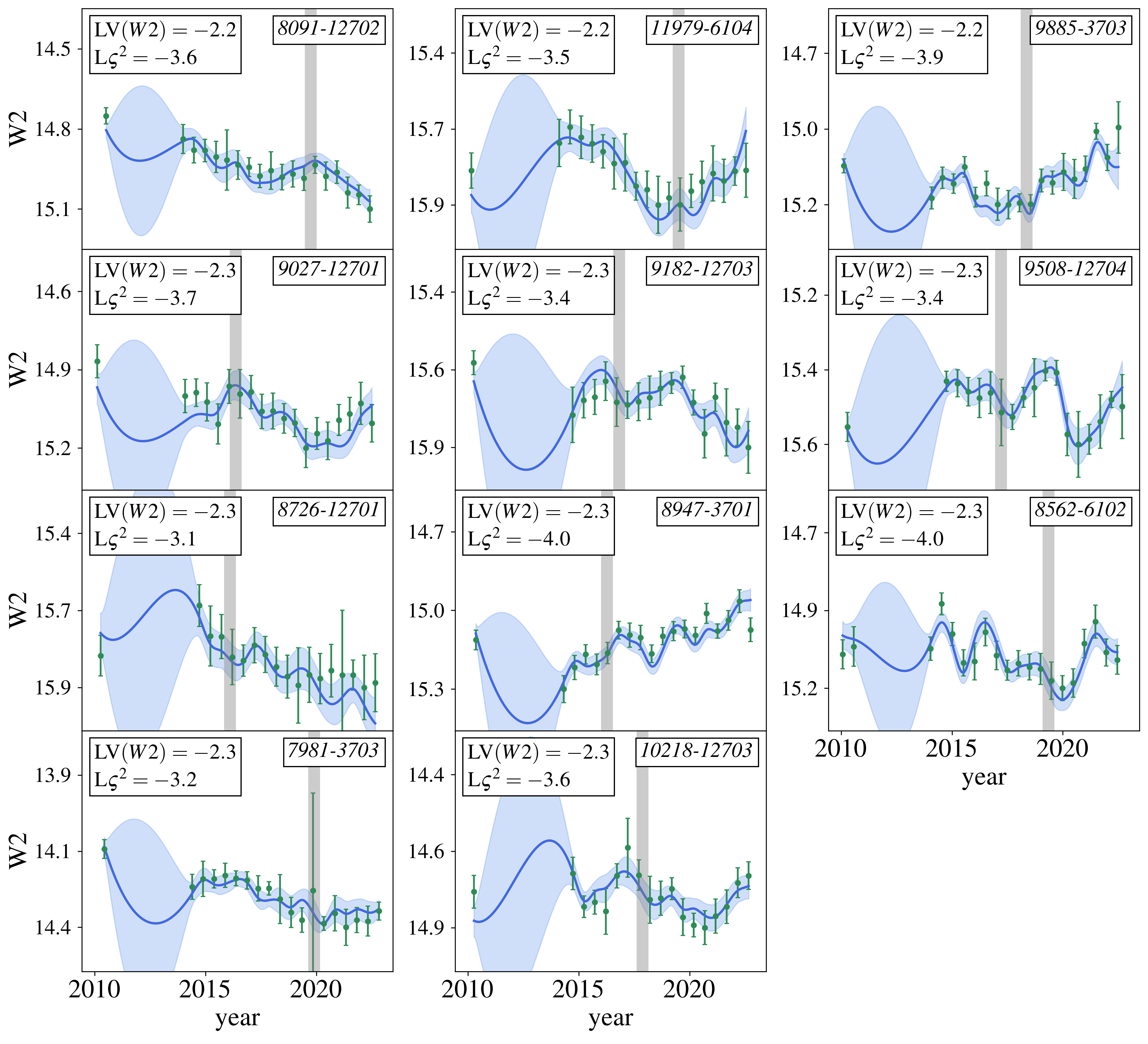}
     \label{fig:lightcurves90}
     %\caption{$W2$ light-curves of mid-IR variable galaxies, ordered by $\log_{10}\Var{(W2)}$ and labelled by MaNGA plate-IFU. The grey line marks when MaNGA observed the galaxy.}
\end{figure*}


\allauthors


\end{document}

%% file: classification.tex
\begin{deluxetable}{ccc}
\tablecaption{Fraction of variable objects from our catalog that are identified as AGN or transients using different classification schemes.\label{tab:classification}}

\tablehead{\colhead{AGN classification scheme} & \colhead{$N/N_{\rm var}$} & \colhead{$N/N_{\rm cat}$} }
\startdata
Broad Line AGN (\citealt{fu23a}) & 0.42 & 0.39  \\
Narrow Line AGN (\citealt{jiyan20a}) & 0.28 & 0.12  \\
Red $W1-W2$ color (\citealt{assef18a}) & 0.29 &  0.44 \\
Red $W1-W2$ color  & 0.46 & 0.39 \\
(our modified criterion, see Sec.~\ref{sec:w12}) & & \\
MIRONGs (\citealt{jiang21a}) & 0.02 & 0.03 \\
\hline
\enddata
\end{deluxetable}

%% file: cat_defs.tex
\startlongtable
\begin{deluxetable}{|c|c|c|c|}

\tablecaption{MANGA-WISE variability catalog key\label{tab:cat_defs} }

\tablehead{\colhead{Column in FITS file}  & \colhead{Quantity} & \colhead{Units} & \colhead{Description}}
\startdata
  {\tt plateifu } & Plateifu & --- & Identifier from MaNGA catalog\\
  \hline
  {\tt ra} & RA & J2000 & Right Ascension of galaxy\\
  \hline
  {\tt dec} & Declination & J2000 & Declination of galaxy\\
  \hline
  {\tt mjd} &  Epoch date & days & Modified Julian Date of each epoch \\
  \hline
  {\tt obs\_per\_ep} & $n$ & --- & Number of observations per epoch\\
  \hline
  {\tt epochs\_count} & $N$ & --- & Number of epochs\\
  \hline
  {\tt mean\_W?\_per\_epoch} & $\overline{W?}$ & $\SI{}{\mag}$ & mean of the W? observations in each epoch \\
  \hline
  {\tt err\_W?\_per\_epoch} & $\sigma_{\rm \overline{W?}}$ &\SI{}{\mag} & error per epoch\\
  \hline
  {\tt expected\_var\_W?\_errs} & $\sigma_{\rm ep}^2$ & 
  \SI{}{\square\mag} & expected variance per epoch \\
  & & &  based on catalog uncertainties\\
  \hline
  {\tt expected\_var\_W?\_mags} & $\Var{(W?_{\rm ep})}$ & 
  \SI{}{\square\mag} & expected variance per epoch \\
  & & & calculated from within-epoch $W?$ magnitudes\\
  \hline
  {\tt expected\_W?\_var\_mean\_errs} & $\sigma^2_j$ & \SI{}{\square\mag}& expected variance in the mean per epoch \\
  & & & based on catalog uncertainties\\
  \hline
   {\tt expected\_W?\_var\_mean\_mags} & $\varsigma^2_j$ & \SI{}{\square\mag}& expected variance in the mean per epoch \\
   & & & based on within-epoch $W?$ magnitudes\\
   \hline
   {\tt observed\_W?\_var} & $\Var{(W?)}$ & \SI{}{\square\mag}& observed variance across all epochs \\
   & & & of the mean magnitudes per epoch\\
   \hline
   {\tt expected\_W?\_var\_all\_errs} & $\sigma^2$ & \SI{}{\square\mag}& expected variance across all epochs \\
   & & & based on catalog uncertainties\\
   \hline
   {\tt expected\_W?\_var\_all\_mags} & $\varsigma^2$ & \SI{}{\square\mag} & expected variance across all epochs based on within epoch \\
   & & &  variances calculated from $W?$ magnitudes \\
\hline
{\tt vi\_flag} & --- & --- & 1 if we believe the WISE variability to be questionable,  \\
 & & & 0 otherwise. We only visually inspected galaxies identified  \\
 & & & as variable by our criteria.\\
\hline
{\tt epoch\_flag} & --- & --- & 1 if the galaxy has 3 or fewer epochs, 0 otherwise\\
\hline
{\tt var\_flag} & --- & --- & 1 if the object is variable in the mid-IR, 0 otherwise\\
\hline
\enddata
\end{deluxetable}

%% file: variability_catalog.tex
\startlongtable
\begin{deluxetable}{ccccccc}
\tablecaption{The MANGA-WISE variability catalog\label{tab:cat1}}
\tablehead{
\colhead{Plateifu} & \colhead{RA} & \colhead{Dec} & \colhead{W1} & \colhead{W2} & \colhead{$\log_{10}\varsigma^2$} & \colhead{$\log_{10}\Var{(W2)}$}  \\
 \colhead{}& \colhead{(deg)} & \colhead{(deg)} & \colhead{(AB mag)} & \colhead{(AB mag)} & \colhead{(AB \SI{}{\square\mag})}& \colhead{(AB \SI{}{\square\mag})}}
 \startdata
8091-6101 & 11.8308 & 14.7035 & 13.62 & 13.85 & -4.058 & -1.239 \\
8091-9101 & 11.8765 & 15.6971 & 14.41 & 15.02 & -3.628 & -1.851 \\
8091-12702 & 12.4139 & 13.6852 & 14.96 & 14.76 & -3.638 & -2.222 \\
12073-3704 & 14.6979 & -1.0971 & 14.78 & 14.73 & -3.82 & -1.254 \\
8093-9102 & 20.8484 & 15.2147 & 13.9 & 14.52 & -3.071 & -1.936 \\
12078-6102 & 29.1378 & -0.175 & 14.7 & 15.38 & -3.306 & -1.831 \\
10223-3701 & 31.5666 & -0.2914 & 13.08 & 13.25 & -3.854 & -0.845 \\
9514-1901 & 33.24 & 14.1028 & 14.58 & 14.76 & -3.601 & -1.856 \\
10222-12701 & 43.0975 & -8.5104 & 13.03 & 13.02 & -3.935 & -1.427 \\
9193-9101 & 46.2942 & -1.0755 & 15.64 & 16.14 & -3.579 & -2.073 \\
9193-12704 & 46.6649 & 0.062 & 15.34 & 15.16 & -3.89 & -1.62 \\
8158-3704 & 61.4533 & -6.3238 & 16.23 & 16.7 & -2.761 & -1.815 \\
8132-6102 & 111.1382 & 41.9506 & 15.21 & 15.92 & -2.993 & -1.668 \\
8132-6101 & 111.7337 & 41.0267 & 15.09 & 15.42 & -3.481 & -1.623 \\
8133-12702 & 113.5109 & 43.5436 & --- & --- & -3.3 & -1.879 \\
8726-12701 & 115.717 & 22.1127 & 14.61 & 15.15 & -3.137 & -2.269 \\
8145-6102 & 116.5535 & 26.923 & 15.58 & 16.05 & -2.585 & -0.823 \\
8710-6102 & 117.9662 & 49.8143 & 13.91 & 14.27 & -3.913 & -0.84 \\
9497-12705 & 118.0743 & 19.5951 & --- & --- & -3.408 & -1.743 \\
8717-1902 & 118.0911 & 34.3266 & 14.97 & 14.32 & -4.05 & -1.716 \\
10216-12704 & 118.1623 & 18.3216 & 14.47 & 14.91 & -3.522 & -1.767 \\
8714-3704 & 118.1842 & 45.9493 & 14.29 & 14.61 & -4.117 & -1.819 \\
10217-3704 & 118.3957 & 14.4733 & 15.05 & 15.62 & -2.75 & -1.921 \\
10218-12703 & 118.6342 & 16.8097 & 13.38 & 13.76 & -3.59 & -2.283 \\
9182-12703 & 118.7745 & 40.0413 & 15.46 & 15.83 & -3.412 & -2.26 \\
8713-9102 & 118.8554 & 39.1861 & 14.61 & 14.7 & -3.891 & -2.02 \\
8718-12701 & 119.1822 & 44.8567 & 15.62 & 16.09 & -3.139 & -1.744 \\
9503-12701 & 119.9728 & 23.3901 & 12.41 & 12.42 & -4.03 & -1.724 \\
9498-3703 & 120.0169 & 23.4378 & 13.99 & 14.42 & -3.848 & -1.858 \\
8940-12702 & 120.0874 & 26.6135 & 12.66 & 12.66 & -4.137 & -1.085 \\
9183-3703 & 121.9208 & 39.0042 & 14.69 & 14.87 & -4.082 & -1.98 \\
9487-3702 & 123.3305 & 46.1472 & 15.31 & 15.71 & -3.525 & -1.291 \\
9487-9102 & 123.8203 & 46.0753 & 13.69 & 13.79 & -4.359 & -2.009 \\
10493-1902 & 124.3274 & 52.0299 & 16.85 & 16.82 & -3.449 & -1.756 \\
8723-3702 & 127.0444 & 55.7124 & 13.51 & 13.96 & -3.82 & -2.15 \\
9508-12704 & 127.106 & 26.3974 & 15.71 & 16.05 & -3.45 & -2.267 \\
8241-9102 & 127.1708 & 17.5814 & 14.34 & 14.47 & -4.174 & -1.789 \\
8725-9102 & 127.1781 & 45.7426 & 15.18 & 15.47 & -3.451 & -1.319 \\
11746-12704 & 127.7849 & 30.8956 & 15.08 & 15.45 & -3.242 & -1.864 \\
9507-12705 & 129.5207 & 25.3295 & 14.36 & 14.88 & -3.326 & -2.155 \\
9502-12703 & 129.5456 & 24.8953 & 13.06 & 13.47 & -3.563 & -1.699 \\
10498-6101 & 129.747 & 26.137 & 13.49 & 13.54 & -3.603 & -1.808 \\
11747-12703 & 130.0098 & 29.8174 & 15.42 & 15.11 & -3.878 & -2.054 \\
8723-6104 & 130.4078 & 54.9186 & 13.96 & 13.81 & -3.701 & -1.449 \\
12511-12705 & 133.9486 & 0.7943 & 13.43 & 13.25 & -4.339 & -2.18 \\
12511-3701 & 133.9762 & 0.8531 & 14.83 & 14.9 & -3.784 & -1.136 \\
12511-3702 & 134.6499 & 1.5304 & 15.11 & 15.44 & -3.802 & -2.068 \\
10500-12703 & 137.1602 & 32.593 & 13.11 & 13.7 & -3.972 & -2.175 \\
8152-9102 & 144.1565 & 34.4774 & 16.21 & 16.79 & -3.424 & -2.106 \\
12512-12702 & 146.3735 & -0.3652 & 14.35 & 14.59 & -3.436 & -2.164 \\
10515-6102 & 147.4028 & 4.1163 & 14.89 & 15.43 & -2.879 & -2.133 \\
10845-12701 & 147.9806 & 3.4834 & 14.87 & 15.2 & -3.625 & -1.882 \\
10516-12701 & 150.5293 & 3.0577 & 13.63 & 14.09 & -4.074 & -1.845 \\
10846-6104 & 154.5278 & 0.0999 & 14.47 & 14.8 & -3.704 & -2.164 \\
11838-12703 & 156.5881 & 0.6849 & 13.37 & 13.07 & -4.422 & -1.318 \\
11865-1901 & 156.5881 & 0.6849 & 13.36 & 13.07 & -4.422 & -1.318 \\
11838-1902 & 156.7275 & -0.5415 & 15.3 & 15.76 & -3.872 & -1.826 \\
10837-3701 & 160.1437 & 0.8174 & 14.28 & 14.75 & -3.454 & -1.969 \\
8999-3702 & 163.4339 & 49.4989 & 14.98 & 15.33 & -3.612 & -2.045 \\
8998-12705 & 163.6638 & 47.8623 & 15.73 & 16.13 & -3.523 & -1.867 \\
8947-3701 & 168.9478 & 50.4016 & 15.51 & 15.21 & -4.043 & -2.269 \\
8466-3704 & 169.5134 & 45.113 & 14.67 & 14.96 & -3.936 & -1.717 \\
9000-1901 & 171.4007 & 54.3826 & 13.79 & 13.75 & -4.575 & -0.898 \\
8992-3702 & 171.6573 & 51.573 & 15.23 & 15.63 & -2.59 & -1.833 \\
8992-12702 & 172.3172 & 51.5232 & 14.02 & 14.39 & -4.005 & -1.341 \\
8990-12705 & 173.5376 & 49.2546 & 16.24 & 16.59 & -3.243 & -1.178 \\
8310-3701 & 179.2945 & 22.2962 & 15.27 & 15.43 & -3.656 & -1.916 \\
8260-9101 & 182.2867 & 44.0032 & 13.95 & 14.24 & -3.785 & -1.309 \\
11833-3702 & 184.0295 & 50.8251 & 13.69 & 14.1 & -4.265 & -1.514 \\
8262-9101 & 184.5438 & 44.4005 & 13.85 & 14.37 & -3.922 & -2.164 \\
12483-9102 & 186.454 & 33.5469 & 14.22 & 14.49 & -3.67 & -1.332 \\
8341-12704 & 189.2133 & 45.6512 & 17.85 & 18.23 & -3.781 & -0.997 \\
11014-9101 & 192.0387 & 26.9392 & 14.19 & 14.46 & -3.992 & -1.635 \\
11863-3703 & 194.0593 & 56.8734 & --- & --- & -3.403 & -2.137 \\
11954-1901 & 196.5874 & 53.3064 & 14.9 & 15.38 & -3.76 & -2.115 \\
8318-3704 & 197.8918 & 44.9331 & 14.48 & 14.91 & -3.821 & -2.109 \\
12514-6103 & 198.2742 & 1.4655 & 17.19 & 17.75 & -3.179 & -2.183 \\
11830-1902 & 198.7493 & 51.2726 & 15.23 & 15.53 & -3.696 & -1.112 \\
8322-3701 & 199.0665 & 30.2645 & 14.5 & 14.88 & -3.799 & -1.835 \\
12514-9102 & 199.2706 & 1.828 & 14.29 & 14.62 & -3.597 & -1.899 \\
8445-3701 & 205.6425 & 35.0203 & 13.64 & 14.2 & -3.533 & -1.994 \\
8446-1901 & 205.7533 & 36.1657 & 15.86 & 15.94 & -3.965 & -1.423 \\
8983-3704 & 206.0079 & 25.9412 & 14.97 & 15.28 & -3.469 & -2.024 \\
8447-9102 & 207.4544 & 40.5374 & 14.53 & 15.0 & -2.669 & -1.707 \\
8334-3703 & 213.2302 & 39.3127 & 15.03 & 15.6 & -3.536 & -2.164 \\
8337-1901 & 214.0964 & 38.191 & 15.28 & 15.59 & -3.27 & -1.946 \\
8333-12705 & 214.2842 & 42.6828 & 16.77 & 17.42 & -2.642 & -1.924 \\
8326-6102 & 215.0179 & 47.1213 & 14.75 & 14.86 & -4.184 & -1.81 \\
8333-3703 & 215.4497 & 42.3279 & 15.66 & 16.14 & -3.14 & -2.074 \\
8547-12701 & 217.63 & 52.7072 & 13.92 & 14.29 & -4.319 & -1.726 \\
11011-1902 & 218.7186 & 48.6619 & 15.42 & 15.63 & -4.131 & -0.897 \\
9867-6104 & 220.8804 & 49.3931 & 13.95 & 14.38 & -4.069 & -2.112 \\
11835-12705 & 220.9555 & 1.926 & 14.24 & 14.71 & -3.331 & -2.215 \\
9024-12705 & 223.8675 & 32.84 & 15.58 & 16.13 & -2.717 & -1.328 \\
8593-12705 & 226.9375 & 51.4528 & 13.61 & 13.92 & -4.293 & -1.566 \\
8596-12702 & 230.1723 & 49.1065 & 13.63 & 13.91 & -4.174 & -1.621 \\
8598-9102 & 230.1723 & 49.1065 & 13.63 & 13.91 & -4.174 & -1.621 \\
9890-1901 & 232.4191 & 30.4859 & 14.91 & 14.93 & -3.817 & -1.82 \\
9870-1902 & 233.1651 & 44.5342 & 17.09 & 17.53 & -3.062 & -0.3 \\
9870-9101 & 233.2834 & 44.5357 & 14.67 & 15.23 & -3.747 & -1.311 \\
8553-1901 & 233.9683 & 57.9026 & 13.74 & 13.48 & -4.599 & -1.945 \\
8488-12702 & 234.3747 & 46.8641 & 15.02 & 15.51 & -3.406 & -2.221 \\
9889-12705 & 235.9988 & 24.9985 & 15.78 & 16.12 & -2.724 & -1.636 \\
11974-12701 & 237.477 & 8.8612 & 15.21 & 15.68 & -2.864 & -1.562 \\
8546-6103 & 237.6516 & 51.398 & 14.96 & 14.89 & -4.191 & -1.699 \\
8564-3701 & 238.8712 & 47.3172 & 15.42 & 15.88 & -3.491 & -2.149 \\
11974-3702 & 239.0804 & 9.8154 & 13.94 & 14.21 & -3.994 & -1.447 \\
9887-1901 & 239.1785 & 29.8132 & 17.85 & 17.83 & -4.311 & -1.694 \\
9094-6101 & 239.4314 & 27.4647 & 14.14 & 14.4 & -3.992 & -1.23 \\
8487-9102 & 240.2142 & 46.4814 & 16.13 & 16.72 & -3.165 & -1.857 \\
8549-12701 & 240.4709 & 45.3519 & 14.49 & 14.97 & -4.172 & -2.202 \\
9032-12701 & 240.4751 & 31.8921 & 14.09 & 14.34 & -3.157 & -1.672 \\
8313-6101 & 240.6581 & 41.2934 & 14.55 & 15.04 & -3.868 & -1.442 \\
9031-12705 & 241.151 & 43.8798 & 15.9 & 16.39 & -3.379 & -1.831 \\
9036-6101 & 241.151 & 43.8798 & 15.86 & 16.36 & -3.379 & -1.831 \\
9885-3703 & 241.1527 & 23.6632 & 14.24 & 14.79 & -3.873 & -2.242 \\
9087-6101 & 241.1933 & 21.9573 & 15.18 & 15.0 & -4.338 & -1.988 \\
8549-12702 & 241.2714 & 45.443 & 14.01 & 14.4 & -4.132 & -1.595 \\
8561-3704 & 241.3254 & 52.1202 & 16.93 & 17.32 & -3.656 & -1.084 \\
9032-12702 & 241.4528 & 30.7171 & 15.76 & 16.29 & -3.07 & -2.004 \\
9090-3701 & 241.7174 & 27.9275 & 15.11 & 15.0 & -4.363 & -1.641 \\
9091-3704 & 241.9447 & 25.5375 & 15.88 & 15.71 & -3.601 & -2.137 \\
9030-12702 & 242.3324 & 30.0108 & 15.61 & 16.01 & -3.435 & -2.063 \\
11944-12704 & 243.2568 & 37.2875 & 14.4 & 14.37 & -3.298 & -1.389 \\
8314-1902 & 243.259 & 39.2371 & 14.06 & 14.53 & -4.025 & -2.184 \\
9046-1902 & 243.555 & 26.0712 & --- & --- & -4.517 & -2.123 \\
9027-12701 & 243.9359 & 31.964 & 13.63 & 14.11 & -3.652 & -2.257 \\
8562-6102 & 244.1027 & 51.9498 & 14.25 & 14.75 & -4.033 & -2.27 \\
8483-12701 & 244.1881 & 48.6003 & 16.51 & 16.82 & -3.309 & -1.592 \\
9048-1902 & 246.256 & 24.2632 & 15.66 & 16.01 & -3.604 & -1.392 \\
11940-12704 & 246.9641 & 36.1556 & 13.89 & 13.85 & -4.313 & -2.102 \\
9892-6102 & 247.4704 & 24.444 & 14.49 & 15.0 & -4.077 & -1.706 \\
11943-6104 & 247.636 & 39.3842 & 13.97 & 14.09 & -4.385 & -1.975 \\
9892-12702 & 247.8147 & 23.8826 & 14.83 & 14.8 & -4.0 & -1.595 \\
9892-3703 & 248.3839 & 24.9847 & 15.19 & 15.52 & -3.469 & -1.931 \\
11757-3702 & 248.4313 & 26.329 & 14.91 & 15.16 & -3.756 & -2.086 \\
11757-12704 & 249.5572 & 26.2818 & 13.86 & 14.32 & -3.798 & -1.982 \\
12667-6104 & 249.9323 & 36.7008 & 16.22 & 16.5 & -3.077 & -2.042 \\
11978-6103 & 250.2818 & 22.8236 & 15.51 & 15.94 & -3.18 & -1.346 \\
8978-6104 & 251.3359 & 42.7578 & 17.2 & 17.67 & -3.363 & -2.195 \\
12651-3703 & 252.1943 & 27.0105 & 15.82 & 16.5 & -2.838 & -2.089 \\
11979-6104 & 253.048 & 22.7565 & 14.56 & 15.17 & -3.492 & -2.229 \\
8612-1901 & 253.4676 & 39.7602 & 12.13 & 12.47 & -4.719 & -1.957 \\
11981-3703 & 254.4339 & 19.3687 & 16.32 & 16.28 & -3.337 & -1.88 \\
11981-6104 & 254.8444 & 20.8298 & 13.86 & 14.35 & -3.754 & -1.479 \\
7959-6103 & 256.2832 & 30.4032 & 14.72 & 15.11 & -4.255 & -2.038 \\
8614-12703 & 258.1185 & 35.8841 & 13.9 & 14.29 & -4.085 & -1.677 \\
7958-9101 & 258.4958 & 33.6071 & 15.53 & 15.87 & -3.371 & -2.07 \\
7991-6104 & 258.8274 & 57.6588 & 14.37 & 14.9 & -3.851 & -1.838 \\
9186-9101 & 260.6664 & 30.8813 & 13.15 & 13.21 & -4.5 & -1.35 \\
9196-12703 & 262.3993 & 54.4944 & 14.42 & 14.53 & -4.599 & -1.888 \\
8624-3702 & 263.0508 & 59.9415 & 13.29 & 13.67 & -4.674 & -1.683 \\
8626-12702 & 263.2054 & 56.9013 & 14.38 & 14.95 & -4.265 & -1.918 \\
7972-3704 & 316.8413 & 11.0664 & 15.99 & 16.2 & -3.023 & -1.378 \\
7815-6104 & 319.1931 & 11.0437 & 14.98 & 14.97 & -3.963 & -1.984 \\
8615-3701 & 321.0079 & -0.3663 & 14.99 & 15.21 & -3.709 & -1.732 \\
8616-12701 & 322.2532 & 0.1826 & 15.75 & 16.34 & -2.961 & -2.138 \\
8619-3701 & 322.9334 & 11.3812 & 16.45 & 16.78 & -3.559 & -1.707 \\
8619-12702 & 323.1159 & 10.1386 & 15.29 & 16.01 & -4.642 & -2.047 \\
12683-6102 & 325.5908 & -8.3649 & 15.43 & 15.86 & -2.996 & -1.946 \\
7977-9101 & 331.1229 & 12.4426 & 13.81 & 14.2 & -4.101 & -1.097 \\
7981-9102 & 338.4101 & 13.2121 & 14.84 & 14.86 & -4.031 & -1.962 \\
7981-3703 & 339.2821 & 14.232 & 13.83 & 14.14 & -3.213 & -2.276 \\
12071-1901 & 345.9218 & 1.0436 & 14.4 & 14.15 & -3.841 & -2.147 \\
12085-6103 & 345.9294 & 13.9265 & 14.06 & 14.52 & -3.828 & -2.194 \\
12071-12702 & 346.3285 & 0.1895 & 12.36 & 12.31 & -4.367 & -2.184 \\
12085-1902 & 346.5688 & 14.429 & 16.33 & 17.02 & -2.989 & -1.971 \\
12071-12703 & 347.3345 & 0.7565 & 16.0 & 16.38 & -3.865 & -1.304 \\
12088-3701 & 348.4187 & 14.021 & 13.37 & 13.48 & -4.05 & -1.918 \\
12089-6104 & 351.8415 & 15.4104 & 13.46 & 13.53 & -4.428 & -1.616 \\
\hline
\enddata
\end{deluxetable}